\documentclass{aa}  

\usepackage{graphicx}
\usepackage{txfonts}
\usepackage{lipsum}
\usepackage{subcaption}
\usepackage{multirow}
\usepackage{url}
\usepackage{upgreek}
\usepackage{lscape}
\usepackage{placeins}         
\usepackage{booktabs}
\usepackage{comment, enumitem, makecell,multirow}
\usepackage{longtable,booktabs,array} 
\usepackage{multirow}
\usepackage{amsmath}
\usepackage[hidelinks]{hyperref}
\makeatletter
\newcommand{\corrmark}{\fnmsep\textsuperscript{\@fnsymbol{1}}}
\renewcommand*\aa@textidlineempty{{\slshape A\&A proofs:}\ manuscript no.~aa59120-26}
\renewcommand*\aa@manuscriptname{%
  manuscript no.~aa59120-26%
  \hspace{\stretch{1}}%
  \copyright ESO \the\year
}
\makeatother

\begin{document}

   \title{Unveiling hidden millihertz quasi-periodic oscillations in 1A~0535+262}

   \author{Zihan Yang\inst{1}
        \and Ruican Ma\inst{2}\fnmsep\thanks{Corresponding authors: Ruican Ma (R.Ma@soton.ac.uk) and Lian Tao (taolian@ihep.ac.cn).}
        \and Mariano M{\'e}ndez\inst{3}
        \and Z.X. Li\inst{1}
        \and Q.C. Zhao\inst{1}
        \and Panping Li\inst{1}
        \and Lian Tao\inst{1}\corrmark
        \and Shuang-Nan Zhang\inst{1}
        \and Liang Zhang\inst{1}
        \and Hua Feng\inst{1}
        \and Xiang Ma\inst{1}
        \and Yue Huang\inst{1}
        }

   \institute{Key Laboratory for Particle Astrophysics, Institute of High Energy Physics, Chinese Academy of Sciences, 19B Yuquan Road, Beijing 100049, China
             \and School of Physics and Astronomy, University of Southampton, Highfield, Southampton, SO17 1BJ, UK
             \and Kapteyn Astronomical Institute, University of Groningen, PO BOX 800, NL-9700 AV Groningen, the Netherlands
            }

   \date{Received 25 January 2026 / Accepted 16 April 2026}

\abstract
{Be/X-ray binary pulsars show transient outbursts and complex timing behaviour, including millihertz quasi-periodic oscillations (QPOs), whose physical origin and energy dependence remain poorly understood. The bright outbursts of 1A~0535+262 in November 2020 provide an ideal laboratory to investigate these properties.}
{We aim to characterise the temporal evolution and energy-dependent properties of the mHz QPO during its 2020 giant outburst, with a focus on trying to detect this feature at lower energies.}
{We use the multi-Lorentzian fitting framework that was recently introduced to jointly model the power spectra and the real and imaginary parts of the cross-spectrum. Our analysis incorporates simultaneous broadband X-ray observations from the \textit{Neutron Star Interior Composition Explorer} (\textit{NICER}) and \textit{Insight}-HXMT, spanning the 0.2--120\,keV energy range.}
{We report the first detection of weak, but significant, mHz QPOs at low X-ray energies ($<27$\,keV), extending their detection to a new energy regime. The centroid frequency evolves from 41 to 93\,mHz, with the peak root-mean-square (rms) amplitude detected in the 50--65\,keV. Throughout the outburst, the QPOs generally exhibit a hard lag between $0.12 {\pi\,{\rm rad}}$ and $0.9 {\pi\,{\rm rad}}$. However, at the outburst peak, the higher-energy bands ($>35$\,keV) display a soft lag of up to $\sim -0.93{\pi\,{\rm rad}}$. We propose that interactions between soft seed photons and an extended outflow located outside the magnetosphere can account for the observed hard lags, although the physical origin of the transient soft lags remains uncertain. Furthermore, we detect a double-peaked mHz QPO only at high energies ($E>35$\,keV) near peak luminosity. The two peaks maintain an approximately constant separation of $2\nu_{\rm spin}$ and exhibit anti-correlated phase evolution. The combination of this constant frequency separation and opposite phase trends is difficult to reconcile within existing theoretical models.}
{Our results indicate that the mHz QPOs in the Be/X-ray binary 1A~0535+262 are closely linked to the coupled evolution of a soft-photon source and a Comptonizing outflow or corona. The joint cross-spectral framework provides a complementary probe of mHz QPOs beyond traditional power-spectral analyses.}


\maketitle
\nolinenumbers

\section{Introduction}
\label{sec:Intro}
Be/X-ray binaries (Be/XRBs) are a subclass of high-mass X-ray binaries consisting of a neutron star (NS) and a Be-type stellar companion. The Be star is typically a rapidly rotating main-sequence or giant star \citep{Reig2011, Rivinius2013},  surrounded by a quasi-Keplerian circumstellar disc, which forms through episodic mass ejections from the stellar equator. This disc plays a key role in enabling mass transfer onto the neutron star and strongly influences the system's X-ray variability \citep{Okazaki2013, Reig2007}.

Most Be/XRBs are transient systems that undergo episodic X-ray outbursts, classified into two types: type I (normal) and type II (giant) \citep{Finger1996GRO,Bildsten1997,Negueruela2001}. 
Type I outbursts are typically periodic, occurring near the periastron passage, and last for a small fraction of the orbital cycle ($\sim$0.2--0.3), with peak luminosities $\lesssim 10^{37}\,\mathrm{erg\,s^{-1}}$. 
Type II outbursts are more luminous ($\gtrsim 10^{38}\,\mathrm{erg\,s^{-1}}$), significantly longer in duration, and do not correlate with a specific orbital phase \citep{Frank2002, Wilson2008}. 
These giant outbursts are often associated with the temporary formation of an accretion disc around the neutron star.

During type II outbursts, Be/XRBs frequently exhibit strong aperiodic flux variability. 
Among this variability, millihertz quasi-periodic oscillations (mHz QPOs) in the $\sim$1--400\,mHz range have drawn particular attention \citep{Bozzo2009, Boroson2000}. 
These QPOs are believed to originate from the inner regions of the accretion flow and serve as important diagnostics of accretion dynamics in strongly magnetised neutron-star systems \citep{Dugair2013, Nandi2012}. 
mHz QPOs have been observed in a variety of sources \citep[e.g.][]{Mukerjee2001,Revnivtsev2001,Devasia2011,Lyu2016,Bult2017} and several models have been proposed, such as the beat–frequency (BF) model \citep{Alpar1985}, the Keplerian–frequency (KF) model \citep{vanKlis1987}, and neutron–star/disc–precession models \citep{StellaVietri1998}. The physical origin of these QPOs, however, remains poorly understood.

Variability in X-ray binaries (XRBs) can be analysed through the power spectrum (PS), which reveals QPOs as narrow features in the Fourier domain \citep{Klis1989, Nowak2000}. 
However, weak or blended QPO components may remain undetected in the PS, particularly when they coexist with strong broad-band noise or overlap with other variability components in the same frequency range \citep[e.g.,][]{Zhang2017, Zhang2020}. 
The cross spectrum (CS) is another widely used tool in timing analysis, providing a quantitative measure of the correlation between light curves in two different energy bands \citep{Vaughan1997, Nowak1999a, Uttley2014}. The CS encodes phase lags at each Fourier frequency and, together with the PS, enables computation of the coherence function, a frequency-dependent measure of the linear correlation between concurrent light curves. Extracting phase lags from the CS typically assumes that the variability component of interest dominates both the PS and the CS within the selected frequency range \citep{Belloni2002, Ingram2019, Alabarta2022}. 
This assumption frequently fails in practice when multiple overlapping components contribute to the variability, leading to ambiguous or inaccurate lag estimates \citep{Wang2021, Wang2022}.

To overcome these limitations, \citet{Mendez2024} introduced a new method that simultaneously fits the PS and the real and imaginary parts of the CS using a multi-Lorentzian model. 
In this framework, each variability component is described by a Lorentzian profile modulated by a phase lag, under the assumption of coherence within each energy band and incoherence among different variability components \citep{Nowak1999a}.
By jointly modelling all variability components in both the PS and CS, this approach allows for the detection of QPO features that would otherwise remain undetectable using the PS alone.
This method also yields reliable estimates of phase lag and coherence, even in the presence of strong contamination from other components. This capability is crucial for uncovering weak or complex QPO signals,  as reported in black hole XRBs such as MAXI J1820+070 \citep{Mendez2024}, Cyg X-1 \citep{Fogantini2025}, and Swift J1727.8--1613 \citep{Jin2025}.

We extend this method to investigate the neutron star XRB 1A~0535+262, a transient X-ray pulsar discovered in 1975 using \textit{Ariel V} \citep{Rosenberg1975, Coe1975}. 
This system consists of a neutron star in a moderately eccentric orbit ($e\sim0.47$) with an orbital period of approximately 111 days and a spin period of $\sim$103\,s \citep{Finger1996}. 
Located at a distance of 2.13\,kpc \citep{BailerJones2018}, 1A~0535+262 exhibits cyclotron resonance scattering features at 45\,keV and 120\,keV, indicating a surface magnetic field strength of $B \sim 4 \times 10^{12}$\,G \citep{Kendziorra1994}.

Since its discovery, 1A~0535+262 has exhibited numerous outbursts. According to the disc truncation model proposed by \citet{Okazaki2001}, the nature of these outbursts depends on the system's orbital eccentricity and disc dynamics \citep{Haigh2004, Coe2006}. 
Among these outbursts, at least ten have been classified as type~II \citep{Camero2012}, during which broad QPOs in the 27–95\,mHz range were observed, notably in 1994, 2009, and 2020 \citep{Finger1996, Camero2012, Ma2022}.

Interestingly, these mHz QPOs have consistently been detected only at relatively high energies. \citet{Camero2012} found that the QPO amplitude increases with photon energy and becomes undetectable below 25\,keV, challenging the explanation of classical BF and KF models. \citet{Ma2022} used broadband observations from \textit{Insight}-HXMT and detected QPOs up to 80\,keV, with the strongest significance in the 50--65\,keV. However, no QPOs were detected below 27\,keV, and the observed energy dependence of the root-mean-square (rms) amplitude appeared inconsistent with current theoretical models.

The absence of low-energy X-ray detections leaves the origin of mHz QPOs unclear, highlighting the need to reassess their behaviour in low-energy bands using the new methodology. In this study, we focus on the 2020 giant outburst of 1A~0535+262, which was the brightest recorded to date, reaching a peak flux of approximately 11~Crab in the \textit{Swift}/BAT band \citep{Mandal2020, Pal2020}. The 2020 outburst presents a unique opportunity to explore the characteristics of mHz QPOs across a broad energy band, as it was observed with high statistical quality by both \textit{Insight}-HXMT (1–250\,keV, 1910\,ks) and the \textit{Neutron Star Interior Composition Explorer} (\textit{NICER}; 0.2–12\,keV, 1900\,ks).

In this work, we apply the new timing analysis method developed by \citet{Mendez2024} to joint \textit{Insight}-HXMT and \textit{NICER} observations of the 2020 outburst of 1A~0535+262, aiming to investigate the broadband (0.2–120\,keV) characteristics of the detected mHz QPO, with particular emphasis on its low-energy behaviour. 
The structure of this paper is as follows: Section~\ref{sec:model} describes the joint fitting model we used. Section~\ref{sec:data_ana} outlines the observations and data reduction procedures, while Section~\ref{sec:result} presents the main results. In Section~\ref{sec:discussion}, we discuss the implications of our findings, and Section~\ref{sec:conclusions} summarises our conclusions.

\section{Mathematical formalism}
\label{sec:model}
We investigate the variability properties of 1A~0535+262 using the new timing method proposed by \citet{Mendez2024}, which fits the PS and both the real and imaginary parts of the CS simultaneously. 
This method enhances the detection of weak or hidden variability components that may not appear significantly in the PS alone but can be recovered through their coherence or phase-lag signatures.

Let $x(t)$ and $y(t)$ denote two simultaneous light curves extracted from different energy bands. 
Their Fourier transforms are given by $X(\nu)$ and $Y(\nu)$, and the corresponding power spectra are defined as:
\begin{equation}
G_{xx}(\nu) = \langle X(\nu) X^*(\nu) \rangle, \quad
G_{yy}(\nu) = \langle Y(\nu) Y^*(\nu) \rangle,
\end{equation}
where $\langle \cdot \rangle$ indicates averaging over data segments.

The complex cross-spectrum is:
\begin{equation}
G_{xy}(\nu) = \langle X(\nu) Y^*(\nu) \rangle = |G_{xy}(\nu)| e^{i \Delta \phi_{xy}(\nu)},
\end{equation}
where $\Delta \phi_{xy}(\nu)$ is the phase lag between $x(t)$ and $y(t)$ at frequency $\nu$.

The intrinsic coherence function is defined as:
\begin{equation}
\gamma_{xy}^2(\nu) = \frac{|G_{xy}(\nu)|^2}{G_{xx}(\nu) G_{yy}(\nu)},
\end{equation}
which quantifies the degree of linear correlation between the signals in different energy bands at each frequency \citep{Vaughan1997,Uttley2014}.

Our joint-fitting approach relies on four assumptions: 
\begin{itemize}[label={\scriptsize$\bullet$}]
\item (i) The variability can be described as a linear combination of Lorentzian components, as commonly adopted in X-ray timing studies \citep[e.g.,][]{Nowak2000,Belloni2002}. 
\item (ii) For each Lorentzian component, the centroid frequency $\nu_{0,i}$ and width $\Delta_i$ are assumed to be identical in all energy bands, such that the profile of the $i$-th component is energy-independent \citep{Mendez2024}. 
\item (iii) Each Lorentzian component is assumed to be perfectly coherent between any two energy bands. 
\item (iv) Any two Lorentzian components that partially overlap in frequency are assumed to be mutually incoherent. This is a widely adopted assumption in modelling neutron-star power spectra using multiple (approximately) independent Lorentzian components \citep[e.g.,][]{vanStraaten2002}.
\end{itemize}

Following assumption (i), we model both power spectra as a sum of $n$ Lorentzian functions:
\begin{equation}
G_{xx}(\nu) = \sum_{i=1}^n A_i L(\nu; \nu_{0,i}, \Delta_i), \quad
G_{yy}(\nu) = \sum_{i=1}^n B_i L(\nu; \nu_{0,i}, \Delta_i),
\end{equation}
where $A_i$ and $B_i$ are the integrated powers in each band, and $L(\nu; \nu_{0,i}, \Delta_i)$ denotes the normalised Lorentzian profile; per assumption (ii), the same $\nu_{0,i}$ and $\Delta_i$ are used for a given Lorentzian in all energy bands.

Following assumption (iii), each Lorentzian is fully coherent between bands $x$ and $y$ (i.e. $\gamma^2_{xy,i}=1$), so the variability in bands $x$ and $y$ is related by a deterministic linear transfer, and the cross-spectrum of the $i$-th component satisfies
\begin{equation}
    |G_{xy,i}(\nu)|^2 = A_i B_i L^2(\nu; \nu_{0,i}, \Delta_i).
\end{equation}

From the above, and using assumption (iv), the total cross-spectrum is expressed as
\begin{equation}
G_{xy}(\nu) = \sum_{i=1}^n C_i L(\nu; \nu_{0,i}, \Delta_i) e^{i \Delta \phi_{xy,i}(\nu)},
\end{equation}
where $C_i = \sqrt{A_i B_i}$ and $\Delta \phi_{xy,i}$ is the phase lag of the $i$-th component.

In this work, we adopt the constant phase-lag model (see \citealt{Mendez2024} for details), where
\begin{equation}
\Delta \phi_{xy,i}(\nu) = 2{\rm \pi} k_i.
\end{equation}

The real and imaginary parts of the CS are then given:
\begin{align}
    \text{Re}[G_{xy}(\nu)] = \sum_{i=1}^n C_i L(\nu; \nu_{0,i}, \Delta_i) \cos(2{\rm \pi} k_i) \\
    \text{Im}[G_{xy}(\nu)] = \sum_{i=1}^n C_i L(\nu; \nu_{0,i}, \Delta_i) \sin(2{\rm \pi} k_i).
\end{align}
The total, frequency-dependent phase lag is reconstructed as:
\begin{equation}
\Delta \phi_{xy}(\nu) = \arg\left( \sum_{i=1}^n C_i L(\nu; \nu_{0,i}, \Delta_i) e^{i \Delta \phi_i} \right).
\end{equation}

\noindent
For clarity, we use $\Delta \phi_{xy}(\nu)$ to refer to the total observed phase-lag, while $\Delta \phi_{\mathrm{QPO}}$ denotes the intrinsic phase lag of an individual QPO component.

The statistical uncertainty of the squared coherence function is given by \citep{Bendat2011, Vaughan1997}:
\begin{equation}
d\gamma^2_{xy} = \frac{\sqrt{2}(1 - \gamma^2_{xy})}{\gamma^2_{xy} \sqrt{N}},
\end{equation}
where $N$ is the number of averaged segments.

This relation shows that the coherence function is substantially more robust to statistical noise than either the PS or CS, especially when $\gamma_{xy}^2 \approx 1$. 
Consequently, components with high intrinsic coherence achieve a higher effective signal-to-noise ratio in the coherence spectrum than in the PS or CS, making coherence-based diagnostics particularly powerful for detecting weak but correlated variability \citep{Fogantini2025, Rout2025}. 
This effect is especially important in broadband-dominated or high-energy regimes, where background noise often suppresses individual spectral features.

The joint modelling of PS and CS not only improves the sensitivity to such weak components, but also allows for a more comprehensive characterisation of their spectral–timing properties, including their energy-dependent coherence and phase-lag behaviour. 

\section{Observations and data analysis}
\label{sec:data_ana}
\subsection{Data reduction}
\subsubsection{\textit{Insight}-HXMT}
\textit{Insight}-HXMT (hereafter HXMT) provides broadband X-ray coverage from 1 to 250\,keV through three onboard instruments: the Low-Energy X-ray Telescope (LE; 1--10\,keV, with an effective area of 384\,cm$^2$; \citealt{Chen2020}), the Medium-Energy X-ray Telescope (ME; 8--35\,keV, 952\,cm$^2$; \citealt{Cao2020}), and the High-Energy X-ray Telescope (HE; 27--250\,keV, 5100\,cm$^2$; \citealt{Liu2020}). 

HXMT monitored 1A~0535+262 with high cadence and high statistics for nearly 50 days during its 2020 giant outburst, collecting a total exposure of over 2.09\,Ms. In this work, we focus on data obtained through the HXMT Core Science Program, spanning the period from 14 November to 24 December 2020 (MJD = Modified Julian Date: 59167.33--59207.72; ObsID P0314316001--P0314316015). Each exposure ID (ExpID) typically provides a long integration time, ranging from 90 to 190\,ks.

Furthermore, to improve the signal-to-noise ratio (SNR), data products from all ExpIDs observed on the same day, typically 5 to 8 per day, were merged into a single daily dataset. The reorganised datasets used in this analysis are summarised in Table~\ref{tab:RegroupHXMT}. 
HXMT data reduction was carried out using {\sc hxmtdas} v2.04. 
Background estimation for the HE, ME, and LE instruments was performed using the corresponding tools—\texttt{HEBKGMAP}, \texttt{MEBKGMAP}, and \texttt{LEBKGMAP}—based on the HXMT background models \citep{Guo2020,Liao2020a, Liao2020b}.

\begin{figure}[htbp]
    \centering
    \includegraphics[width=0.48\textwidth]{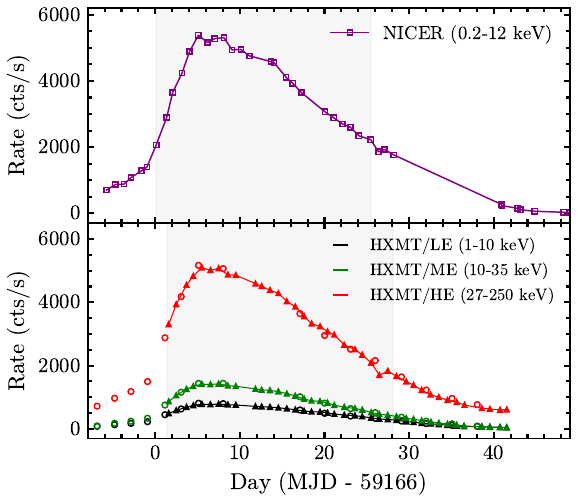}
    \caption{Long-term X-ray light curves of the 2020 outburst of 1A~0535+262. The top panel shows the \textit{NICER} count rate in the 0.2--12\,keV band, with each observation represented by a purple square. The bottom panel displays the \textit{Insight}-HXMT light curves from the Core Science Program (P0314316). The LE (1--10\,keV), ME (10--35\,keV), and HE (27--250\,keV) count rates are represented by the black (bottom), green (middle), and red (top) triangles, respectively. Additional observations from proposal P0304099 (PI: P. Reig) are marked with circular symbols. Each data point represents a single observational epoch. The shaded grey region denotes the interval during which the mHz QPO was most significantly detected at energies above 27\,keV \citep{Ma2022} and selected for cross-spectral analysis.}
    \label{LC}
\end{figure}

\begin{figure*}[htbp]
    \centering
    \begin{minipage}[c]{12cm}
        \centering
        \includegraphics[width=\linewidth]{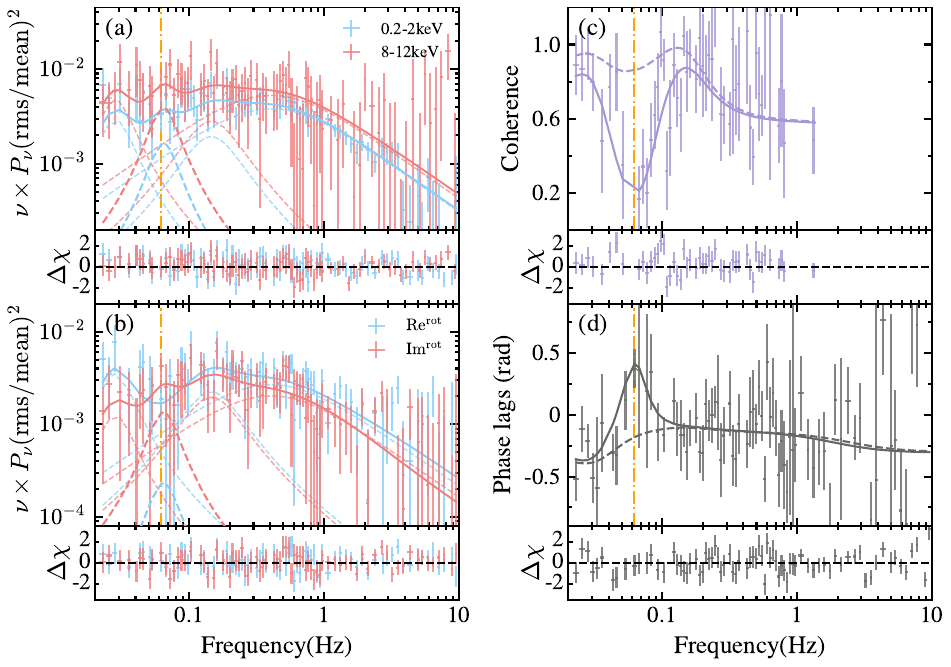}
    \end{minipage}\hfill
    \begin{minipage}[c]{5.5cm}
        \caption{Joint fit to the \textit{NICER} data of 1A~0535+262 from Day~1. (a) PS of reference (0.2--2\,keV; blue) and subject (8--12\,keV; red) bands. (b) $45^\circ$-rotated cross spectrum: $\mathrm{Re}^{\rm rot}$ (blue) and $\mathrm{Im}^{\rm rot}$ (red; QPO $\mathrm{Re}^{\rm rot}$ sign-flipped, see Fig.~\ref{fig:linear_nicer}). In (a) and (b), solid/dashed curves show the total 5-Lorentzian model and individual components. (c) Coherence and (d) phase lags, with models including (solid) and excluding (dashed) the QPO. Yellow dash-dotted line marks the QPO centroid frequency, coinciding with a coherence dip and phase-lag rise. Data above $\gtrsim 1$\,Hz are logarithmically rebinned with a factor of $\exp(1/10)$. Bottom sub-panels show residuals $\Delta\chi=(\mathrm{data}-\mathrm{model})/\mathrm{error}$.}
        \label{Nicer_Example}
    \end{minipage}
\end{figure*}

\begin{figure*}[htbp]
    \centering
    \begin{minipage}[c]{12cm}
        \centering
        \includegraphics[width=\linewidth]{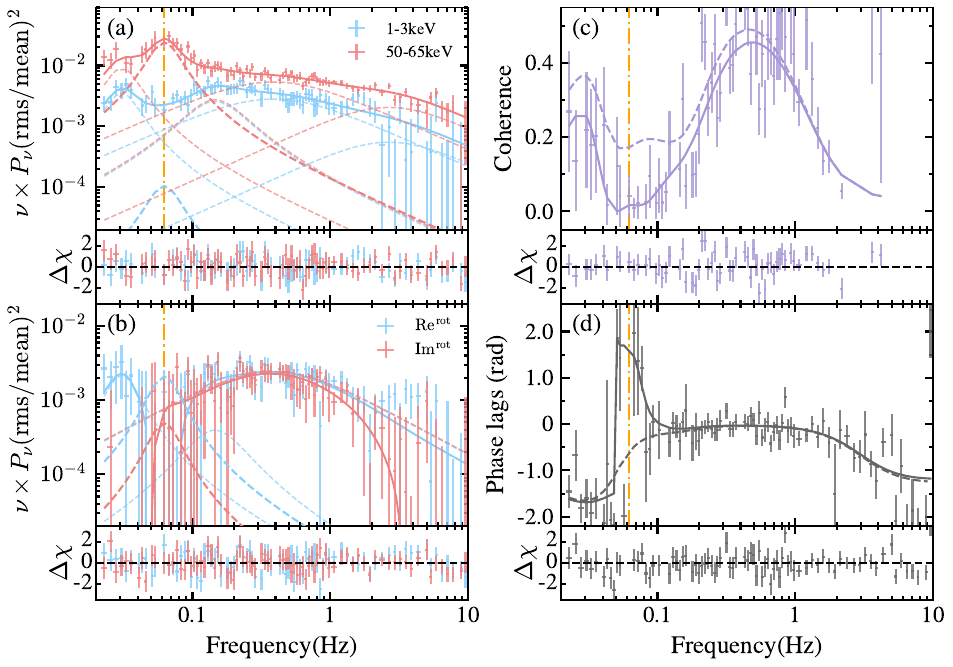}
    \end{minipage}\hfill
    \begin{minipage}[c]{5.5cm}
        \caption{Same layout as Fig.~\ref{Nicer_Example}, but for HXMT data. The reference and subject bands are 1--3 and 50--65\,keV, respectively. Negative Lorentzian contributions to $\mathrm{Re}^{\rm rot}$ or $\mathrm{Im}^{\rm rot}$ are plotted with reversed sign; see Fig.~\ref{fig:linear_5C}.}
        \label{HE_Example}
    \end{minipage}
\end{figure*}

\subsubsection{\textit{NICER}}
To explore data in the lower energy band, we incorporated observations from the \textit{Neutron Star Interior Composition Explorer} \citep[\textit{NICER;}][]{Gendreau2016}, mounted on the International Space Station. \textit{NICER} is well-suited for this purpose due to its lower energy coverage (0.2--12\,keV) and large effective area ($\geq$2000\,${\rm cm}^2$). The \textit{NICER} observations of 1A~0535+262 span from 11 November 2020 (MJD 59160.23) to 25 February 2021 (MJD 59263.01).

Data reduction was carried out using \texttt{HEASoft} v6.31.1 and the \texttt{NICERDAS} pipeline (version 2022-12-16 V010a), along with the latest calibration files (\texttt{CALDB vxti20221001}). Cleaned event files were produced with \texttt{nicerl2}.
Since we focus on the bright phase of the outburst, the background contribution was negligible compared to the source count rate and was therefore not taken into account in our analysis.

The \textit{NICER} and HXMT light curves of the 2020 outburst of 1A 0535+262 are presented in Fig.~\ref{LC}. The shaded regions indicate the observation periods analysed in this study; detailed information corresponding to these intervals is provided in Table~\ref{tab:RegroupHXMT}.

\begin{table}[htbp]
\caption{Energy bands for cross-spectral analysis.}
\label{tab:energy_bands}
\centering
\small
\setlength{\tabcolsep}{3pt}
\renewcommand{\arraystretch}{1.1}
\begin{tabular}{@{}lccc@{}}
\toprule
Telescope & \makecell{Reference\\band (keV)} & \makecell{Subject\\band (keV)} \\
\midrule
\multirow{3}{*}{\textit{NICER}} & \multirow{3}{*}{0.2--2} & 2--5 \\
& & 5--8 \\
& & 8--12 \\
\midrule
\multirow{2}{*}{LE/\textit{Insight}-HXMT} & \multirow{2}{*}{1--3} & 3--5 \\
& & 5--10 \\
\midrule
ME/\textit{Insight}-HXMT & 1--3 & 10--35 \\
\midrule
\multirow{6}{*}{HE/\textit{Insight}-HXMT} & \multirow{6}{*}{1--3} & 27--35 \\
& & 35--40 \\
& & 40--50 \\
& & 50--65 \\
& & 65--80 \\
& & 80--120 \\
\bottomrule
\end{tabular}
\end{table}

\subsection{Computation of PS, CS, coherence, and phase lag}
To compute the PS, we employed the \texttt{GHATS} \footnote{\url{https://github.com/ghats-timing/ghats-timing.github.io}} software to perform Fast Fourier Transform (FFT) using a segment length of 393.216 seconds and a time resolution of 6\,ms. The adopted $\Delta t$ places the LE, ME, and HE light curves on a common time grid for the cross-correlation analysis, while the chosen segment length of 393.216\,s (i.e. $2^{16}$ time bins at 6\,ms resolution) preserves the low-frequency resolution required for the mHz-QPO analysis. This configuration yields a minimum frequency (and frequency resolution) of 0.002543\,Hz, and a Nyquist frequency of 83\,Hz. The light curves were not background-subtracted prior to the Fourier analysis; instead, we accounted for the background statistically in the fractional-rms normalisation of the power spectra, which mitigates the dilution of intrinsic variability by background counts, particularly at high HXMT energies \citep{Belloni1990}. To account for the Poisson noise, we subtracted its level from the 60--80\,Hz band. We then logarithmically rebinned the PS, increasing the bin width by a factor of $\approx 1.047$ ($10^{1/50}$) for each successive bin. To study the energy dependence of the variability, we divided the \textit{NICER} data into four energy sub-bands: 0.2--2, 2--5, 5--8, and 8--12\,keV. For the HXMT data, we adopted ten sub-bands \citep[see also][]{Ma2022}: 1--3, 3--5, 5--10\,keV for LE, 10--35\,keV for ME, and 27--35, 35--40, 40--50, 50--65, 65--80, 80--120\,keV for HE. 

The CS was computed using the same segment length and time resolution as the PS. We selected the 0.2--2\,keV band as the reference band for \textit{NICER}, and the 1--3\,keV energy band for HXMT data. The corresponding reference and subject bands are listed in Table~\ref{tab:energy_bands}. To correct for the partial correlation of photons present simultaneously in narrow and broad energy bands, we subtracted the average real part of the CS over the same 60--80\,Hz range used for noise correction in the PS \citep{Mendez2024, Bellavita2025}. 
Phase lags and coherence were derived from the PS and CS (see Section~\ref{sec:model} and \citealt{Mendez2024} for further details). We found that in most observations, the phase lags remain close to zero across a broad frequency range, implying that the imaginary part of the CS is substantially smaller than the real part. To improve the robustness of the model fitting, we applied a $45^\circ$ rotation to the cross vectors of the two energy bands in the Fourier plane. This transformation equalizes the contributions of the real and imaginary components without altering the best-fitting parameters \citep[see][]{Mendez2024}.

\subsection{Joint fitting model}
\label{subsec:fittingmodel}
To independently measure each variability component, we fit the PS in two bands, $G_{xx} (\nu)$ and $G_{yy} ({\nu})$, as well as the real and imaginary parts of the CS, $\mathrm{Re}[G_{xy} (\nu)]$ and $\mathrm{Im}[G_{xy} (\nu)]$ (for more details, see Section~\ref{sec:model} and \citet{Mendez2024}). Since both the phase lag,  $\Delta\phi(\nu)$,  and coherence $\gamma_{xy}(\nu)$ can be derived from the above four independent spectra, we have flexibility in selecting a subset of four out of the six available spectra for joint modelling \citep{Fogantini2025, Rout2025}. In our analysis, we choose to fit $G_{yy}(\nu)$ (subject band PS), $\mathrm{Re}[G_{xy}(\nu)]$, $\mathrm{Im}[G_{xy}(\nu)]$, and $\gamma_{xy}(\nu)$. 
This selection is motivated by the observation that $\gamma_{xy}(\nu)$ exhibits prominent structure in our dataset, and it serves as a sensitive diagnostic, especially for identifying variability components that exhibit weaker PS and/or CS. 
For the MJD~59167 observation, we utilised the phase-lag spectra instead of the coherence spectra, as the latter were of insufficient quality for this dataset; the same choice was adopted for all subject bands with $E>80$\,keV.

We perform the joint fit using \texttt{XSPEC} v12.14.0h over the 0.02--10\,Hz frequency range. During the fitting process, the centroid frequencies ($\nu_0$) and FWHM ($\Delta$) of the Lorentzians were linked across the spectra but allowed to vary freely, while the normalisations in the PS and CS were allowed to vary independently. Parameter uncertainties were estimated in \texttt{XSPEC} using the \texttt{error} command with $\Delta\chi^2=2.706$, corresponding to the 90\% confidence interval for one parameter of interest. 
Following \citet{Mendez2024}, a Lorentzian is considered significant if the normalisation of that component in any of the PS, or the corresponding quantity in the CS, divided by its negative $1\sigma$ error, exceeds 3. A component is retained in the joint-fitting model if it is significant in at least one PS or in the CS.
From the best-fit model, both the PS of the reference band and the phase lags were derived.\footnote{To exclude the contribution of the pulse and its harmonics of 1A~0535+262, we excluded the corresponding contaminated frequency bins (up to five times the frequency), following the approach described in \citet{Ma2022}.}

\section{Results}
\label{sec:result}
\subsection{Hidden mHz QPOs in the low-energy band}
\label{sec:fit_example}
We first analysed the coherence and phase lags across all HXMT observations, with emphasis on the low-energy band (LE; <27\,keV). In 1A~0535+262, during the outburst peak (MJD 59167--59194, Days 1--28; ObsID P0314316001--P0314316011), the coherence shows a marked decrease and the phase lags exhibit a pronounced feature at the QPO frequencies identified in the high-energy PS (Fig.~\ref{Cohlag_noFit}), spanning the full energy range.  As noted in Appendix~\ref{app:ledata}, similar features appear in some LE data but at lower significance than in ME/HE; during the rise and decay phases of the outburst, limited statistics prevent a firm confirmation of such structures. 
Similar behaviour has been reported in several BHXBs (e.g., MAXI~J1820+070, \citeauthor{Bellavita2025}~\citeyear{Bellavita2025}; Cyg~X-1, \citeauthor{Fogantini2025}~\citeyear{Fogantini2025}; and Swift~J1727.8--1613, \citeauthor{Jin2025}~\citeyear{Jin2025}), and has been attributed to the presence of a hidden QPO-like component.

To verify the characteristics of the coherence and phase lags, particularly in the low-energy band, we analysed the set of \textit{NICER} data, taking advantage of its larger effective area compared with LE. 
The \textit{NICER} observations reproduce the main features seen in the HXMT data, especially during the outburst peak (MJD 59167--59191, Days 1--25; ObsID 3200360130--3200360152). 

Given \textit{NICER}'s higher statistical quality (clearer structure in coherence and phase lags) and better low-energy coverage, we use the \textit{NICER} (0.2--12\,keV) in place of the LE (1--10\,keV) data; for the high-energy band, we continue to use ME and HE data (10--120\,keV). To quantitatively characterise the structures in the coherence and phase-lag spectra and explore the low-energy properties of 1A~0535+262, we apply the joint fitting method proposed by \citet[][see also Section~\ref{subsec:fittingmodel} in this work]{Mendez2024}.

We present representative low-energy fitting results using \textit{NICER} data in Fig.~\ref{Nicer_Example}, including the PS of the reference (0.2--2\,keV) and subject (8--12\,keV) bands, the real and imaginary parts of the CS (after a $45^\circ$ rotation), the coherence, and the phase-lag spectra. The data are well described by a model with five Lorentzian components, yielding $\chi^2 = 403.92$ for 399 degrees of freedom ($dof.$) One of these Lorentzians, centred at $61.9 \pm 4$\,mHz (vertical dash-dot line), represents an additional feature not identifiable from the PS alone. To check its robustness, we repeated the fitting without this extra component. 
In the coherence and phase-lag panels, the model derived from this reduced fit (dashed curves) leaves systematic residuals: it fails to reproduce the observed coherence dip at $\sim$62\,mHz and the associated bump in the phase-lag spectrum. This demonstrates that the additional Lorentzian is required by the data, with a detection significance of $3.2\sigma$ in the joint-fit model.
The centroid frequency of this feature is consistent with the QPO frequency measured in the high-energy PS (see below). At the same frequency, the coherence drops sharply from $\sim$0.9 to $\sim$0.2, accompanied by a bump-like feature in the phase-lag spectrum peaking at $\sim$0.4\,rad. Although this variable component does not produce a distinct peak in the PS at low energies, the corresponding structure in the coherence and phase lag provides strong evidence for an underlying variability component analogous to the mHz QPOs observed at higher energies. We therefore suggest this feature is the ``hidden'' mHz QPO: a component not directly visible in the PS, but required to reproduce the observed coherence and phase-lag behaviour.

When the subject band is HE (50--65\,keV), the PS of the hard band exhibits a prominent and narrow QPO peak, as reported in \citet{Ma2022}. 
An illustrative fit is shown in Fig.~\ref{HE_Example}. 
At the QPO centroid frequency, the phase-lag spectrum displays a steep rise (the ``phase-lag cliff''), while the coherence function shows a localized dip, consistent with the behaviour observed in the lower-energy bands. For comparison, we also computed model predictions after removing the QPO component from the fit. In this case, the model failed to reproduce the PS in the 50--65\,keV, confirming the $12.4\sigma$ detection significance of the QPO component. The corresponding coherence and lag curves are shown as dashed lines in Fig.~\ref{HE_Example}. At low frequencies ($f \lesssim 0.2$\,Hz), the coherence remains low ($\lesssim 0.4$) and is only weakly affected. In contrast, the phase-lag spectrum becomes nearly flat, failing to reproduce the upturn observed at the QPO frequency. These results further indicate that the QPO properties hidden in the lower-energy bands are essentially the same as those observed in the high-energy bands.

\subsection{Energy dependence of mHz QPOs}
\begin{figure}[htbp]
    \centering
    \includegraphics[width=0.40\textwidth]{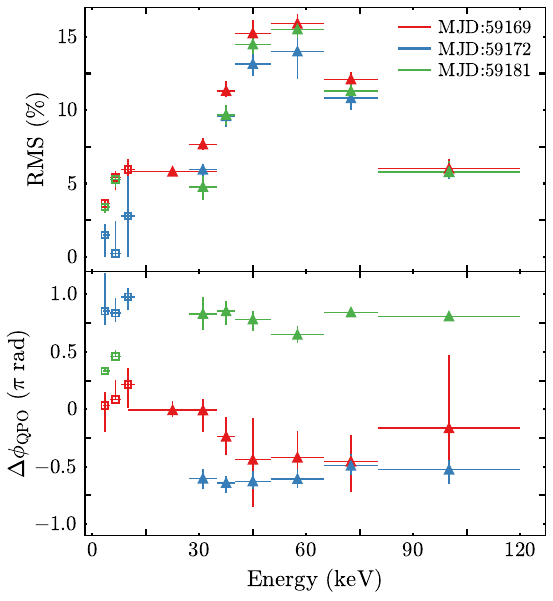}
    \caption{Energy dependence of the mHz QPO properties in 1A~0535+262 for three epochs (MJD~59169, 59172, and 59181; corresponding to Days~3, 6, and 15). Top panel: fractional rms amplitude as a function of energy. Bottom panel: energy-resolved phase lag, $\Delta\phi_{\rm QPO}$, measured with respect to the reference band. Triangles and squares denote HXMT and \textit{NICER} measurements, respectively. Colours indicate the outburst phase: red (rise), blue (peak), and green (decay). In each energy band, the QPO was modelled with a single Lorentzian component.}
    \label{QPO_Ene}
\end{figure}

Figure~\ref{QPO_Ene} presents the energy-dependent evolution of the QPO properties, including fractional rms amplitude and phase lag ($\Delta\phi_{\rm QPO}$), across three representative epochs corresponding to the rise, peak, and decay phases of the outburst. 

The rms amplitude exhibits a pronounced energy dependence. In the low-energy band ($\textless$12\,keV), the fractional rms remains below $\sim$5\%. Above this energy, the amplitude rises with energy, peaking at $\sim$15--18\% around 60\,keV, and subsequently decreases to $\sim$6\% at higher energies.

The phase lag, $\Delta\phi_{\rm QPO}$ (Fig.~\ref{QPO_Ene}, right panel), shows a pronounced and phase-dependent evolution with energy. At soft X-ray energies ($\lesssim12$\,keV), the lag differs markedly among epochs: it is close to zero on Day~3 (MJD~59169; $\sim0.03{\pi\,{\rm rad}}$--$0.22{\pi\,{\rm rad}}$), remains around $\sim0.4{\pi\,{\rm rad}}$ on Day~13 (MJD~59181), and reaches the largest values on Day~6 (MJD~59172; $\sim0.84{\pi\,{\rm rad}}$--$0.98{\pi\,{\rm rad}}$). With increasing energy, Day~3 stays near zero up to $\sim31$\,keV, then turns negative and reaches a minimum of $\sim-0.46{\pi\,{\rm rad}}$ at $\sim72.5$\,keV, followed by a partial recovery to $\sim-0.17{\pi\,{\rm rad}}$ at $\sim100$\,keV (with larger uncertainties at the highest energies). On Day~6, the lag drops rapidly from $\sim{\pi\,{\rm rad}}$ at $\lesssim12$\,keV to negative values, reaching $\sim-0.64{\pi\,{\rm rad}}$ at $\sim35$--$40$\,keV and remaining at $\sim-0.5{\pi\,{\rm rad}}$ above $\sim70$\,keV. In contrast, Day~13 stays positive across the full band, peaking at $\sim0.78{\pi\,{\rm rad}}$--$0.86{\pi\,{\rm rad}}$ around $\sim27$--$40$\,keV, showing a shallow dip to $\sim0.65{\pi\,{\rm rad}}$ near $\sim60$\,keV, and stabilizing at $\sim0.84{\pi\,{\rm rad}}$ at higher energies.

\begin{figure*}[htbp]
\centering
\begin{minipage}[t]{\textwidth}
\centering
\includegraphics[width=0.82\textwidth]{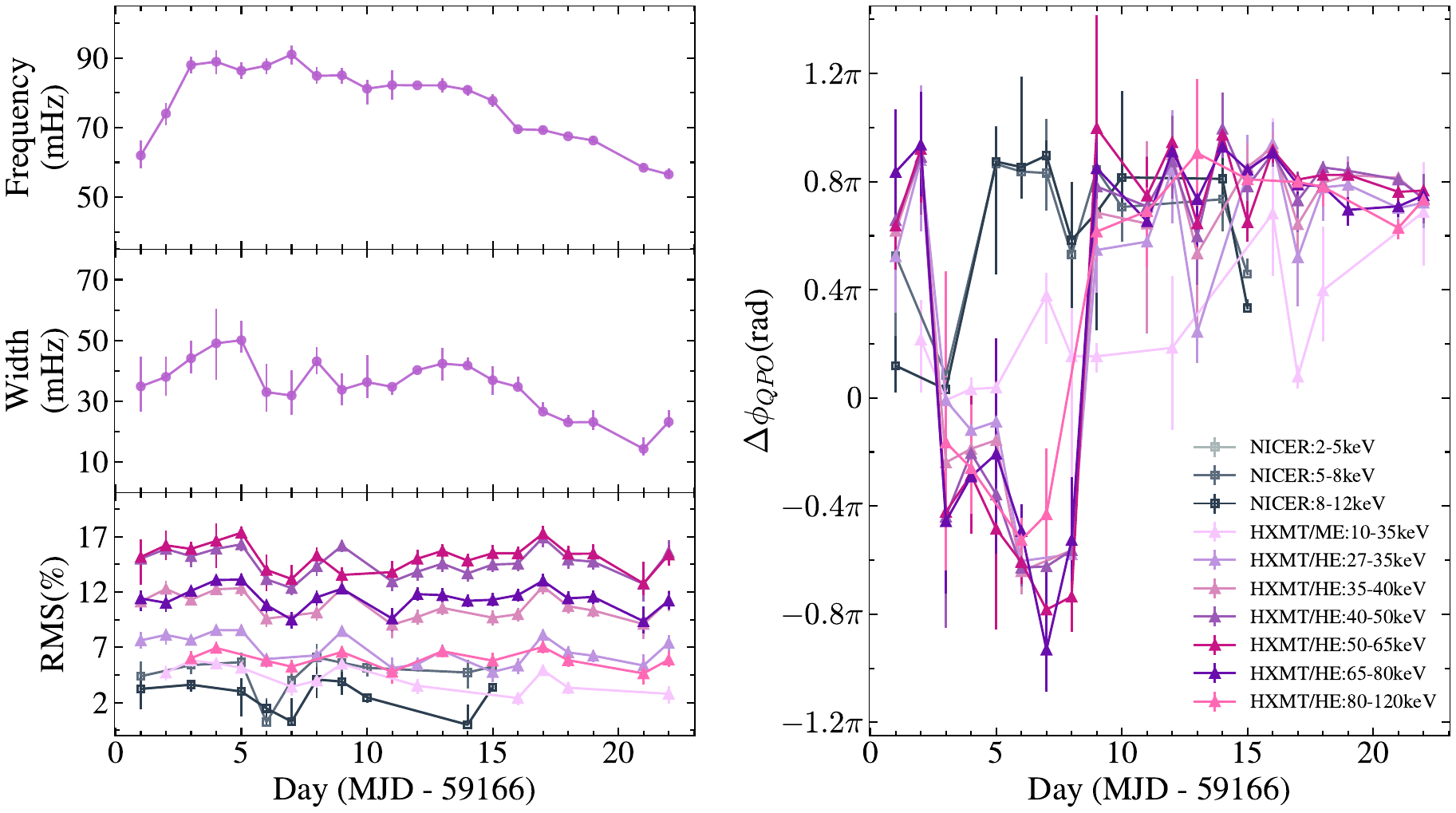}
    \caption{Time evolution of mHz QPO parameters in 1A~0535+262 during the 2020 outburst. The left three panels present the centroid frequency, HWHM, and rms amplitude of the QPOs as functions of MJD. The right panel shows the QPO phase lag ($\Delta\phi_{\rm QPO}$) for the energy bands listed in the legend, measured relative to 0.2--2\,keV for \textit{NICER} and 1--3\,keV for HXMT; \textit{NICER} data are shown in grey and HXMT data in violet. The QPO signal in each energy band was fitted with a single Lorentzian component.}
\label{QPO_MJD}
\end{minipage}

\vspace{2em} 

\begin{minipage}[t]{\textwidth}
\centering
\includegraphics[width=0.98\textwidth]{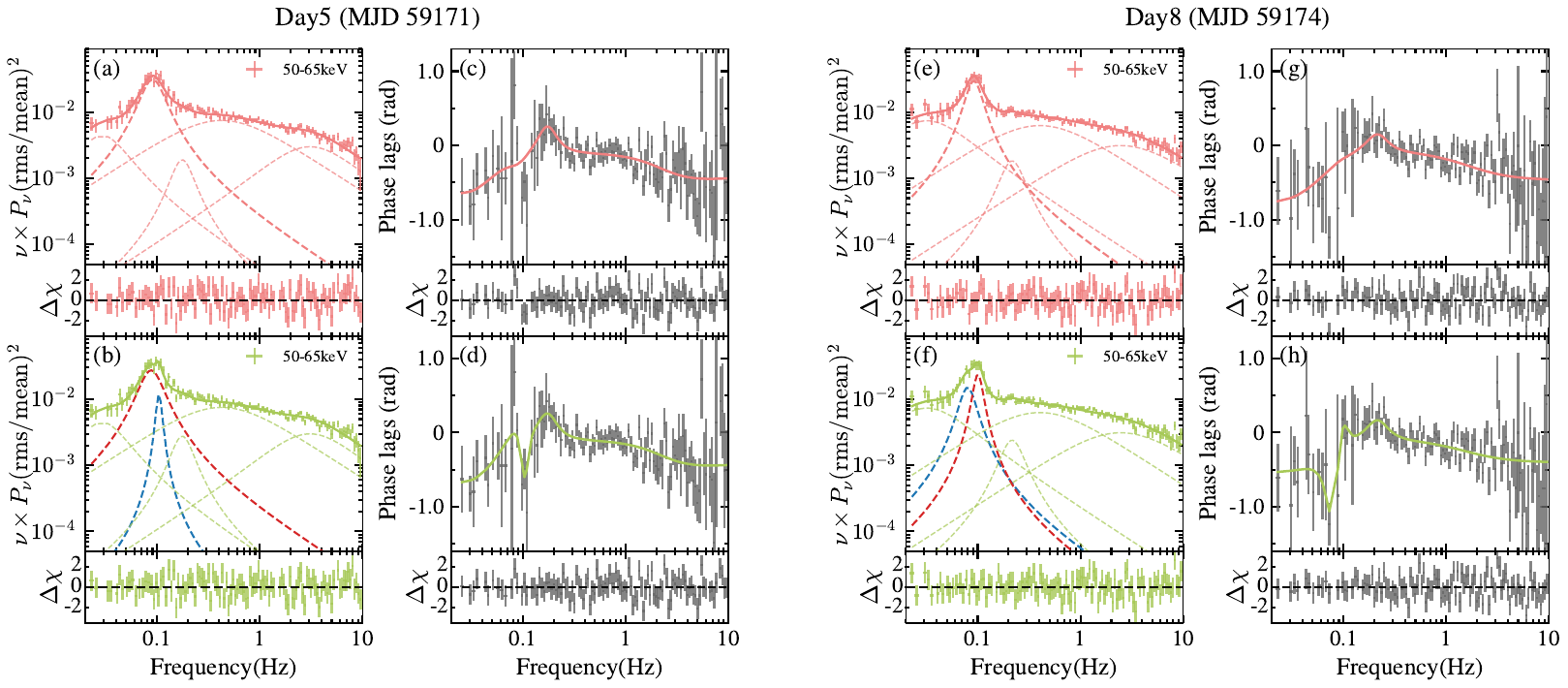}
\caption{Single- and double-QPO fits to the HXMT/HE 50--65\,keV subject band on Day~5 (MJD~59171; panels a--d) and Day~8 (MJD~59174; panels e--h), shown for the power spectrum $\nu P_{\nu}$ and the corresponding phase-lag spectrum relative to the HXMT/LE 1--3\,keV reference band. Panels a, c, e, and g show the single-QPO model (five Lorentzians; red solid curves), whereas panels b, d, f, and h show the double-QPO model (six Lorentzians; green solid curves). Dashed curves denote individual Lorentzian components; in the double-QPO model, the LF and HF QPO peaks are highlighted with red- and blue-dashed curves, respectively, consistent with Fig.~\ref{Double_Ene}. Residuals ($\Delta\chi$) are shown beneath each spectrum.}
\label{fit_DoublePeak}
\end{minipage}
\end{figure*}

\begin{figure}[htbp]
    \centering
    \includegraphics[width=0.48\textwidth]{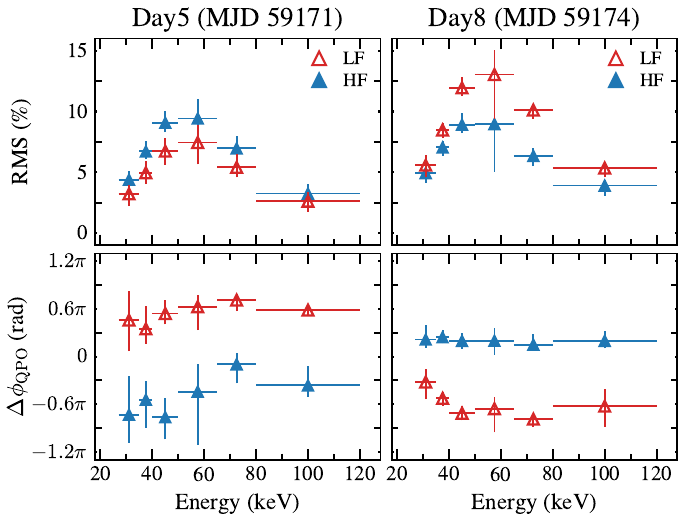}
    \caption{Energy dependence of the double-peak QPO properties in 1A~0535+262 on MJD~59171 and 59174. Top panels: RMS amplitude as a function of energy. Bottom panels: Phase lags ($\Delta\phi_{\rm QPO}$). The filled blue and open red triangles represent the higher- (HF) and lower-frequency (LF) QPOs, respectively. The differences in both rms amplitude ($\sim 4.5\%$ on Day~5 and $\sim 6.6\%$ on Day~8) and phase lag ($\sim0.9\pi$\,rad) remain approximately constant across the 27--120\,keV energy range.}
    \label{Double_Ene}
\end{figure}

\subsection{Time evolution of the mHz QPO}
To explore the time-dependent behaviour of the mHz QPO, we divide the analysis into two energy regimes: soft bands ($<12$\,keV; shown in bluish tones) and hard bands ($>12$\,keV; shown in violet hues) in Fig.~\ref{QPO_MJD}. The left panels show the evolution of the QPO centroid frequency, half-width at half maximum (HWHM), and fractional rms amplitude, while the right panel presents the phase lag, $\Delta\phi_{\rm QPO}$, over the outburst.

The QPO centroid frequency rises rapidly from $\sim62$\,mHz to a maximum of $\sim91$\,mHz during the early outburst (Days 1--7; MJD~59167--59173), and subsequently decreases gradually to $\sim56$\,mHz in the late decay. This evolution broadly follows the luminosity profile inferred from the X-ray light curve \citep[see also][]{Ma2022}. The HWHM shows moderate fluctuations (typically $\sim23$--$50$\,mHz) without a clear systematic trend, although it tends to be narrower in the late decay ($\lesssim25$\,mHz). The fractional rms displays a strong energy dependence but only weak temporal variability: it stays at a few per cent in the \textit{NICER} soft bands and increases to $\sim(6$--$17)\%$ in the HXMT hard bands, with no clear systematic trend with outburst phase.

The phase-lag evolution shows a clear contrast between soft and hard energies (right panel of Fig.~\ref{QPO_MJD}). In the \textit{NICER} soft bands ($<12$\,keV), $\Delta\phi_{\rm QPO}$ is predominantly positive and is typically around $\sim0.8{\pi\,\rm rad}$ during most of the rise and the peak epoch, with a brief drop to values consistent with zero around Day~$\sim3$ and a decrease toward the end of the \textit{NICER} coverage (down to $\lesssim0.5{\pi\,\rm rad}$). The HXMT/ME 10--35\,keV lag is generally smaller (typically $\lesssim0.4{\pi\,\rm rad}$) during the rise/peak, but becomes larger in the decay, reaching $\sim(0.6$--$0.7){\pi\,\rm rad}$ at late times, albeit with sizeable uncertainties.

In contrast, the HXMT/HE bands exhibit a pronounced sign reversal during the peak epoch. The lags are positive at the onset (Days $\sim1$--2; $\sim(0.6$--$0.9){\pi\,\rm rad}$), then turn negative during Days $\sim3$--8, reaching the most negative values around Day $\sim7$--8 (ranging from $\sim -0.56\pi\,\rm rad$ to $-0.93\pi$\,rad, depending on energy). After Day $\sim9$, the lags flip back to positive and remain approximately stable at $\sim(0.7$--$0.8){\pi\,\rm rad}$ through the decay, with only modest energy-to-energy scatter.

\subsection{The double-peaked QPO profile}
\label{sec:2peak}
During the peak of the outburst, a double-peaked structure appears near the QPO centroid frequency in the high-energy ($>27$\,keV) power spectra. This feature is not significantly detected at lower energies, consistent with the findings of \citet{Ma2022}. In this section, we re-examine this double-peaked profile and extend the analysis to the associated phase-lag behaviour for two representative epochs, Day~5 and Day~8 (MJD~59171 and 59174).

Figure~\ref{fit_DoublePeak} presents the joint modelling of the PS and the corresponding phase-lag spectra in the HXMT/HE 50--65\,keV band for both epochs, using HXMT/LE 1--3\,keV as the reference band. We tested two alternative descriptions of the QPO feature: a single-QPO model composed of five Lorentzians (red) and a double-QPO model with six Lorentzians (green), where the additional component accounts for the resolved double peak. To ensure a direct comparison, the frequencies and widths of all non-QPO components were tied between the single- and double-QPO models.

On Day~5, the single-QPO fit gives $\nu_{\rm QPO}=86.7 \pm 3.1$\,mHz. In the double-QPO model, this feature is resolved into two components, hereafter referred to as the lower-frequency (LF) and higher-frequency (HF) peaks. Their centroid frequencies are $\nu_{\rm LF}=83.5^{+3.9}_{-4.2}$\,mHz ($Q_{\rm LF}\simeq1.7$) and $\nu_{\rm HF}=103.9^{+4.2}_{-3.5}$\,mHz ($Q_{\rm HF}\simeq7.8$), separated by $\sim0.02$\,Hz, where $Q\equiv \nu/\Delta$. A similar decomposition is found on Day~8: the single-QPO centroid is $\nu_{\rm QPO}=89.8^{+2.9}_{-3.2}$\,mHz, whereas the two-component fit yields $\nu_{\rm LF}=78.0 \pm 7.6$\,mHz ($Q_{\rm LF}\simeq2.5$) and $\nu_{\rm HF}=99.7^{+3.6}_{-5.8}$\,mHz ($Q_{\rm HF}\simeq4.3$). The implied frequency ratio, $\nu_{\rm HF}/\nu_{\rm LF}\approx1.25$, disfavours a harmonic origin (i.e., a 1:2 fundamental-harmonic relationship).

While both models provide comparable representations of the overall PS, the double-QPO model more accurately reproduces the sharp phase-lag transition at the QPO frequencies, which appears as a narrow, structured feature in the lag spectrum. Importantly, such a two-component description is not required at lower energies for the same epochs. For example, the phase-lag spectra in HXMT/ME 10--35\,keV (reference: 1--3\,keV) and in \textit{NICER} 5--8\,keV (reference: 0.2--2\,keV) are adequately described by a single broad QPO component.

The energy dependence of the two resolved components is summarised in Fig.~\ref{Double_Ene}. On Day~5, both LF and HF rms amplitudes show a convex trend with energy, peaking around 50--65\,keV, with the HF component generally being stronger. The phase lags are clearly component-dependent: the LF peak remains positive (ranging from $\sim0.35\pi\,\rm rad$ to $0.71\pi\,\rm rad$) with only mild energy dependence, whereas the HF peak stays negative and becomes slightly less negative at higher energies (typically between $\sim-0.76\pi\,\rm rad$ and $-0.09\pi\,\rm rad$). On Day~8, the overall rms level increases, and the LF component becomes dominant above $\sim40$\,keV. Meanwhile, the lag signs reverse: the LF lags become negative (down to $\sim-0.79\pi\,\rm rad$), while the HF lags remain mildly positive and nearly energy-independent ($\sim0.24\pi\,\rm rad$). 

In summary, the double-peaked description is required to reproduce the high-energy lag structure but is not necessary for the contemporaneous lower-energy lag spectra. Moreover, the two resolved peaks exhibit distinct phase lags, yet maintain an approximately constant phase separation of $\sim\pi\,\rm rad$ across the hard X-ray band.

\section{Discussion}
\label{sec:discussion}
Using high-statistics, broadband (0.2--120\,keV) observations from \textit{NICER} and HXMT, we apply the new analysis technique proposed by \citet{Mendez2024}, originally developed for black-hole X-ray binaries and here employed for the first time in a neutron star X-ray binary. With this new method, we detect 47--98\,mHz QPOs in 1A 0535+262 during its 2020 outburst. While such QPOs had previously been reported only in the hard X-ray band \citep[e.g.,][]{Finger1996, Camero2012, Ma2022}, our analysis provides the first significant detection of these features in the soft X-ray band, at energies below 27\,keV.
We further extend the analysis of the double-peaked QPOs observed at the outburst peak \citep[see also][]{Ma2022} by investigating their phase lags using the same method of \citet{Mendez2024}. 
Across the energy bands, the two QPO components exhibit an almost constant phase-lag offset of $\sim1.1{\pi\,{\rm rad}}$. 
Temporally, their phase evolution is anti-correlated, with the relative offset transitioning from $\pi\,{\rm rad}$ to $-\pi\,{\rm rad}$ around Day~6, indicating a full $2\pi\,{\rm rad}$ reversal.
Finally, we combine the newly uncovered soft X-ray characteristics of the mHz QPO with insights from the double-peaked structure to explore the possible physical origin of these oscillations.

\subsection{Detection of low-energy mHz QPOs and their physical implications}
We find that the mHz QPO is detected not only in the harder X-ray band ($>27$\,keV) but also, as a weaker yet significant feature, in the soft band ($<27$\,keV). 
This conclusion is supported by the joint PS+CS fitting analysis, in which the inclusion of a QPO component substantially improves the fit to both the coherence and phase-lag spectra, even though no distinct peak is visible in the soft-band PS (see Fig.~\ref{Nicer_Example}). 
The coherence shows a localized dip and the phase lag exhibits a bump at the QPO frequency, confirming the presence of a correlated variability component. 
This demonstrates that the modulation is not confined to the hard X-ray emitting radiation but is also linked to soft-emission components, such as the magnetosphere–inner disc interface. 

\subsubsection{Review of BF and KF scenarios}
Two models are commonly invoked to explain mHz QPOs in neutron-star X-ray binaries (NSXBs). In the beat-frequency (BF) model, accretion material at the inner disc edge orbits at the Keplerian frequency and is accreted onto the neutron star under magnetic control; the stellar magnetic field, rotating at the neutron-star spin frequency, modulates both accretion rate and luminosity \citep{Alpar1985, Shaham1987, vdKlis2006}. In the Keplerian-frequency (KF) model, QPOs arise from partial obscuration by disc inhomogeneities near the inner edge \citep{vanKlis1987}. Earlier non-detections of QPOs below $\sim27$\,keV \citep[e.g.][]{Finger1996, Camero2012, Ma2022} were taken as evidence against both models \citep[][]{Ma2022}. Here, we re-examine these scenarios in light of our new detections of the QPO in the softer X-ray band.

Assuming that the QPO frequency reflects the dynamical timescale at the truncated inner disc, $r_k(\nu_K)=({GM}/{4{\pi}^2\nu_K^2})^{1/3}$, adopting a canonical neutron star with $M=1.4\,M_\odot$ and using the spin period of this system $P_s\simeq103$\,s ($\nu_s\simeq9.7$\,mHz), the observed QPO centroid frequency $\nu_{\rm QPO}\simeq41$--$93$\,mHz implies, for the KF model ($\nu_K=\nu_{\rm QPO}$), $r_k\simeq (1.41\text{--}0.82)\times10^{9}\ \mathrm{cm}$, and for the BF model ($\nu_K=\nu_{\rm QPO}+\nu_s$), $r_k\simeq (1.22\text{--}0.76)\times10^{9}\ \mathrm{cm}$.
To compare with the magnetospheric radius, we compute the standard Alfv\'en radius, $r_A \simeq 3.2\times10^{8}\ \mathrm{cm}\ ({M}/{1.4\,M_\odot})^{-1/7} R_6^{10/7} B_{12}^{4/7} L_{37}^{-2/7}$ and set $r_M=\xi r_A$ with $\xi\simeq0.5$--$1$. For 1A~0535+262, using representative parameters, $B_{12}\simeq4.7$, $R_6=1$, $L_{37}\simeq1$--$12$ \citep{Kong2021}, we obtain $r_M\simeq (0.19\text{--}0.78)\times10^{9}\ \mathrm{cm}$, comparable to the radii inferred from the QPO frequency, $r_k\simeq(0.76\text{--}1.41)\times10^{9}$\,cm (KF and BF ranges combined). This is consistent with the interpretation that the mHz QPO originates from variability at the magnetosphere–disc boundary. Moreover, the observed $\nu_{\rm QPO}$--$L_X$ relation broadly follows the $\nu_{\rm QPO} \propto L_X^{3/7}$ scaling expected in standard BF/KF frameworks \citep{Finger1996,Ma2022} (for HXMT results in 2020, see Fig.~6 in \citet{Ma2022}).

The correlation of QPO frequencies with the inner disc radius, together with the detection of low-energy QPOs, is consistent with both the KF and BF models. However, the rms–energy dependence requires further consideration: the rms of the QPO peaks at intermediate energies (50--65\,keV) and decreases toward both the soft ($\lesssim27$\,keV) and hard ($\gtrsim65$\,keV) bands. As reported by \citet{Ma2022}, this behaviour is not accounted for by either model. The KF model does not predict energy-dependent rms, with the variations arising solely from partial disc obscuration, while the bulk and thermal Comptonization model of the accretion column (\texttt{bwcycl}; \citealt{Becker2007, Ferrigno2009}) likewise does not reproduce such an energy dependence \citep[see][]{Ma2022}.

The suppression of rms variability at $\lesssim$2--3\,keV can be attributed to dilution by stable disc emission \citep[e.g.,][]{Palombara2006, Uttley2014}. However, this explanation alone cannot account for the gradual decline extending up to $\sim$50\,keV in 1A~0535+262, where thermal Comptonization is expected to dominate \citep[e.g.,][]{Becker2007, Ferrigno2009}. Moreover, the reduction of rms at higher energies ($\gtrsim$65\,keV) cannot be simply explained by photon up-scattering, as this process can enhance the rms in some black-hole X-ray binaries \citep[e.g.,][]{Yu2024}. This may indicate that, for the highest-energy photons, the scattering/transfer process becomes less effective at imprinting the QPO modulation, leading to reduced rms. In summary, the observed symmetric rms–energy profile—with a peak at 50–65\,keV and a decline toward both lower and higher energies—suggests a more complex origin of the energy-dependent variability than previously thought.

\subsubsection{Phase-lag evolution with luminosity and beam geometry}
\label{sec:plag_qpos}
The QPO phase lags in 1A~0535+262 show a strong dependence on both photon energy and source luminosity. 
In the soft band ($E<27$\,keV), the lags remain hard, increasing from $\sim0.1\pi$\,rad during the rise to $\sim0.9\pi$\,rad close to the luminosity maximum, and then decreasing to $\sim0.4\pi$\,rad during the decay. 
Conversely, in the hard band ($E\gtrsim35$\,keV) the lag undergoes a transient reversal near the outburst peak, evolving from $\sim0.8\pi$\,rad in the rise to $\sim-(0.8$--$0.9)\pi$\,rad at maximum luminosity, before returning to $\sim0.7$--$0.8{\pi\,\rm rad}$ in the decay. 
Similar lag reversals have been observed in black-hole binaries (e.g., XTE~J1550$-$564, GRS~1915$+$105) and are associated to structural changes in the Comptonizing region \citep{Wijnands1999,Cui2000,Pahari2013}.

\begin{figure}[htbp]
    \centering
    \includegraphics[width=0.45\textwidth]{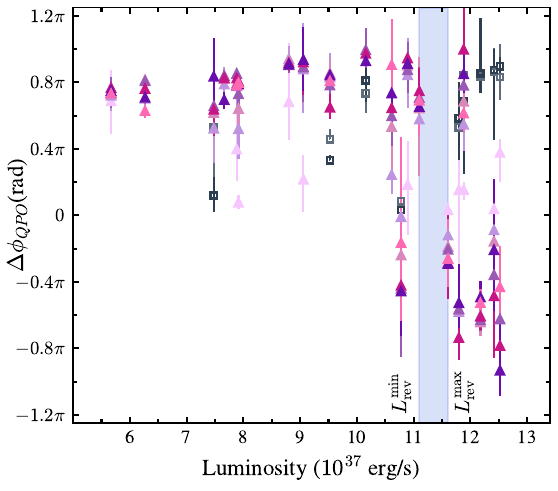}
    \caption{Luminosity–phase–lag relation of the mHz QPO in 1A~0535+262. $\Delta\phi_{\rm QPO}$ in different energy bands (colours as in Fig.~\ref{QPO_MJD}) is shown versus bolometric luminosity. The light-blue shaded region indicates the transition range, $L_{\rm rev} \sim 1.11-1.16 \times10^{38}$. A transient soft lag appears only for $L\gtrsim L_{\rm rev}$. All QPOs were fitted with a single-component model.}
    \label{Lumi_lag}
\end{figure}

We define the lag-reversal luminosity as $L_{\rm rev}\simeq(1.11$--$1.16)\times10^{38}\,\mathrm{erg\,s^{-1}}$ (vertical shaded region in Fig.~\ref{Lumi_lag}).
Below this threshold ($L\lesssim L_{\rm rev}$), the lags are hard across all bands, typically clustering around $0.6\pi$--$0.9\pi$\,rad, with occasional excursions down to $\sim0.1\pi$\,rad and up to $\sim1.1\pi$\,rad. Within a narrow luminosity interval around the outburst peak, $L\simeq(1.16$--$1.25)\times10^{38}\,\mathrm{erg\,s^{-1}}$, the high-energy lags ($E\gtrsim35$\,keV) become negative, reaching $\Delta\phi_{\rm QPO}\approx-(0.6$--$0.9)\pi$\,rad, whereas the low-energy lags remain positive. As the luminosity falls below $L_{\rm rev}$, the soft lags disappear and all bands revert to hard lags, demonstrating that the lag behaviour is tightly regulated by luminosity.

We first evaluate whether the observed lags could arise from local geometric effects, such as energy-dependent beaming and/or multiple emission components (e.g. hotspots or accretion columns; \citealt{Caballero2011,Mushtukov2018}). If these lags were produced locally within the emission region, the intrinsic formation and scattering timescales for soft and hard photons near the surface/column—on the order of $\sim 10^{-5}$ s \citep{Mushtukov2015}—would be far shorter than the observed seconds-scale time lags. 
In addition, hard-photon production via bulk Comptonization within the magnetosphere is unlikely to be efficient given the limited scattering depth \citep[e.g.][]{Becker2007}. 
Moreover, although local radiation patterns can be highly anisotropic \citep[e.g.][]{Poutanen2003,Poutanen2006}, viewing-angle effects are typically coupled to the neutron-star spin frequency rather than producing a stable phase offset at the QPO frequency, unless the QPO itself is driven by a unique geometric modulation (e.g., warped jet precession in MAXI J1820+070; \citealt{MaX2021}).
We therefore suggest that local mechanisms may contribute, but they are insufficient to account for the large-amplitude, luminosity-dependent lag evolution.

A stable phase lag at the QPO frequency is more naturally explained if it arises from delays introduced by scattering and/or irradiation/reprocessing between spatially distinct regions \citep[e.g.][]{Kotov2001,Arevalo2006,Karpouzas2020}. We therefore consider a global scenario in which the lags arise from the coupled variability of the accretion column and an extended hot Comptonizing medium outside the magnetosphere (hereafter, the outflow/corona). 
The measured QPO time lags span approximately $-5$\,s to $+8$\,s. A simple order-of-magnitude estimate based on the light-travel distance, $c|\Delta t|$, yields a characteristic scale of $\sim(1.5$--$2.4)\times10^{11}$\,cm, which is orders of magnitude larger than the magnetospheric radius ($R_{\rm m}\sim10^{8}$--$10^{9}$\,cm), yet remains below the neutron-star Roche-lobe radius ($R_{\rm L}\sim 10^{12}$\,cm). 
Although $c|\Delta t|$ should be interpreted as an upper limit, the large mismatch between $c|\Delta t|$ and $R_{\rm m}$ suggests that the dominant lag-producing region cannot be confined within the magnetosphere. Instead, a more plausible interpretation is that the QPO originates at larger radii in the outer accretion disc, where scattering and irradiation/reprocessing can naturally generate energy-dependent QPO lags.

Within this framework, we propose that the lag evolution is linked to the interaction of soft photons with the outflow/corona. The coupling between the inner disc and the neutron-star magnetic field may drive an extended outflow through magnetic and radiative forces, and dissipation of magnetic energy can maintain a hot, low-density Comptonizing medium at large radii \citep{Romanova2009, Lii2012}. The presence of an outflow is also qualitatively consistent with radio detections \citep{Eijnde2022}, which may trace synchrotron emission associated with open field lines and/or out-flowing plasma in the transition region. At $L\lesssim L_{\rm rev}$, hard lags can be produced if soft seed photons from the inner disc and/or magnetospheric inflow are Compton up-scattered in the hot outflow/corona.

However, the transient soft lags at high luminosity are not straightforward to explain within this picture. One possibility is that near the outburst peak, the outflow becomes increasingly optically thick while remaining hot, shielding a substantial fraction of the soft seed photons. At the same time, hard X-rays from the Comptonizing medium may irradiate the disc and/or dense material in the inner flow more efficiently and be reprocessed into softer emission. If this reprocessed component is viewed under a favourable geometry, it may dominate the soft band and thus produce the observed soft lag. The strength of this effect remains uncertain, and a quantitative explanation of the observed soft lags will require further theoretical constraints and more detailed modelling.

\subsection{Insight from the double-peaked QPO profile at the outburst peak}
\label{sec:double_peak}
During the outburst peak (MJD~59171–59177), we detect a double-peaked mHz QPO exclusively in the hard energy band ($E \gtrsim 35$\,keV). This feature is evident not only in the PS \cite[see also][]{Ma2022}, but is further supported by the corresponding lag behaviour. 
The separation between the centroid frequencies of the two peaks is $\simeq0.02$\,Hz, which is consistent with being twice the spin frequency of 1A~0535+262. The double peak appears only at the outburst maximum and is temporally coincident with the soft-lag window, implying that its geometry is luminosity dependent.

In Sec.~\ref{sec:plag_qpos}, we discussed a scenario where soft seed photons originate near the neutron star, while hard photons are predominantly produced at larger radii within an extended, hot Comptonizing medium created by a strong, magnetically-dissipative outflow outside the magnetosphere.

At high luminosities, numerical simulations suggest that accretion columns may inflate and transition toward a fan-beam dominated emission pattern \citep[e.g.][]{ZhangLZ2022,Sheng2023}, which would enhance the irradiation of the disc/outflow. In addition, the accretion flow may undergo a regime transition from radiation-inefficient to radiation-dominated state, further modifying the reprocessing/scattering environment. These structural changes allow a larger fraction of seed photons to be emitted at wider angles, facilitating more efficient illumination of the outflow/corona and boosting the hard-band flux. However, these geometric considerations alone do not account for the observed double-peaked QPO. Although vertically stratified \citep[e.g.][]{Basko1976,Becker2007,West2017} or hollow-column structures \citep[e.g.][]{ZhangLZ2025} may account for multi-component variability at high luminosities, the fixed frequency separation $\simeq 2\nu_{\rm spin}$ and the distinct phase-lag evolution of the two peaks impose strong constraints on such interpretations. Consequently, these localized models appear insufficient to fully reconcile the observed properties of the double-peaked QPO, suggesting that a more complex coupling between the pulsation geometry, the disc, and the global outflow dynamics is required.

\section{Conclusions}
\label{sec:conclusions}
Based on simultaneous \textit{NICER} and HXMT observations of the Be/X-ray binary 1A~0535+262 in the 0.2--120\,keV band during the 2020 giant outburst, we applied the recently proposed combined power and cross spectrum multi-Lorentzian fitting method \citep{Mendez2024} to neutron star X-ray binaries for the first time, enabling a systematic study and characterisation of the broadband timing properties of mHz QPOs.
Our main conclusions are as follows:

\begin{itemize}[label={\scriptsize$\bullet$}]

\item We detected 41--93\,mHz QPOs in 1A~0535+262 not only in the hard X-ray band, as previously reported \citep[][]{Finger1996, Camero2012, Ma2022}, but also, for the first time, identified a hidden mHz QPO in the soft X-ray band ($<27$\,keV) using the new timing analysis method proposed by \citet{Mendez2024}. 

\item The QPO centroid frequency shows a positive correlation with luminosity. The QPO rms exhibits a convex dependence on energy, peaking at $\sim$50--65\,keV ($\sim17$\%) and decreases both at lower and higher energies, reaching values as low as $\sim4$\% in the softest band.

\item For luminosities below the lag-reversal threshold ($L\lesssim L_{\rm rev}$), the QPO shows hard phase lags at all energies, ranging from $0.4{\pi\,{\rm rad}}$ to $0.9{\pi\,{\rm rad}}$. Near or above $L_{\rm rev}$, the phase lags in the hard band transition to soft lags, reaching $\sim -0.93{\pi\,{\rm rad}}$, while the soft energies maintain hard lags. We discuss that interactions between soft seed photons and a strong, extended outflow may naturally introduce seconds-scale hard lags, whereas the origin of the transient soft lags remains uncertain.

\item In the hard band, we confirm double-peaked QPOs with a constant separation of $\Delta\nu\simeq0.02$\,Hz. The two components exhibit distinct phase lags with opposite signs, maintaining a nearly constant phase separation of $\sim\pi$\,rad. This anti-phase coupling is difficult to reconcile with existing models.
\end{itemize}

These results highlight the power of energy-resolved, coherence-based timing techniques in revealing weak/hidden variability in accreting X-ray pulsars. However, the origin of mHz QPOs, especially the double-peaked QPOs, remains uncertain, warranting further theoretical investigation in the future.

\begin{acknowledgements}
This work made use of data from the Insight-HXMT mission, a project funded by the China National Space Administration (CNSA) and the Chinese Academy of Sciences (CAS), and of observations from the NICER mission supported by NASA. We sincerely thank Sergey Tsygankov for useful discussions and suggestions. We also thank the referee for the constructive comments that improved the quality of this paper. This work was supported by the National Key R\&D Program of China (Grant No. 2021YFA0718500). Additional support was provided by the National Natural Science Foundation of China (Grant Nos. 12122306, 12025301, 12333007, and 12103027), the Strategic Priority Research Program of the Chinese Academy of Sciences, and the China's Space Origins Exploration Program.
\end{acknowledgements}

\bibliographystyle{aa} 
\bibliography{Ref}

\begin{appendix}
\onecolumn
\section{Hidden mHz QPOs in LE data}
\label{app:ledata}
\begin{figure*}[htbp]
    \centering
    \begin{minipage}[c]{12cm}
        \vspace{0pt}
        \centering
        \includegraphics[width=\linewidth]{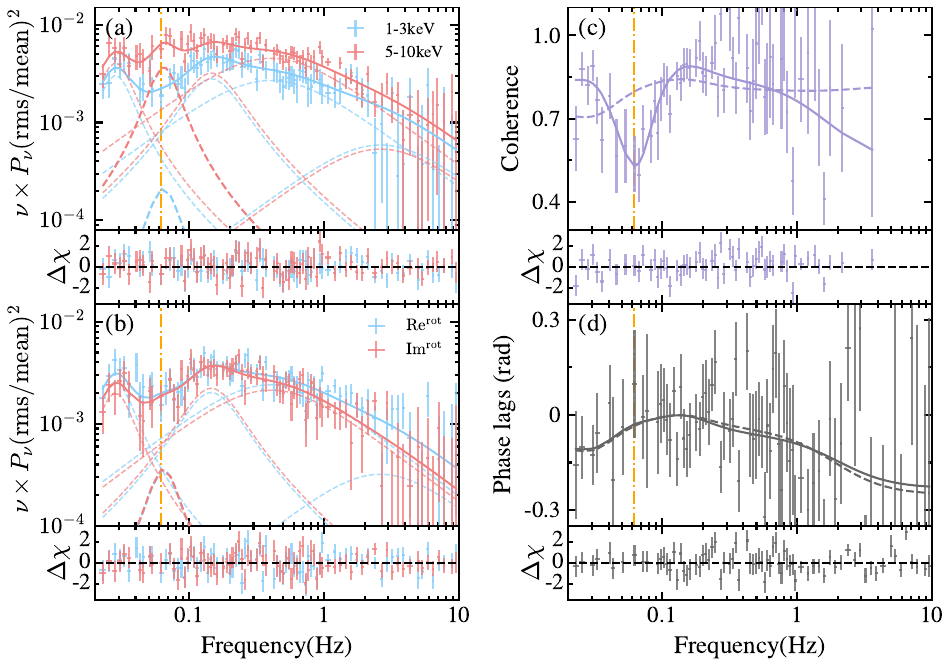}
    \end{minipage}\hfill
    \begin{minipage}[c]{5.5cm}
        \vspace{0pt}
        \caption{Same as Fig.~\ref{Nicer_Example}, but using \textit{Insight}-HXMT data. The selected energy bands are 1--3\,keV for the reference band and 5--10\,keV for the subject band.}
        \label{LE_Example}
    \end{minipage}
\end{figure*}

To verify our results, we applied the same multi-Lorentzian fitting procedure described above to the HXMT/LE dataset. Since most LE data do not exhibit significant features in the coherence or lag (see also Section~\ref{sec:fit_example}). Here, we select the best-quality dataset (Day 1; see Fig.~\ref{LE_Example}) to demonstrate the presence of weak mHz QPOs in the LE data.

We also present the CS of Fig.~\ref{Nicer_Example} and Fig.~\ref{HE_Example} on a linear y-axis, where the contributions of the Lorentzian components to the CS, or the data themselves, can be negative. Such details were not displayed in the main text (Sec.~\ref{sec:fit_example}).
\begin{figure*}[htbp]
    \centering
    \begin{minipage}[c]{12cm}
        \vspace{0pt}
        \begin{subfigure}[t]{0.49\linewidth}
            \vspace{0pt}
            \centering
            \includegraphics[width=\linewidth]{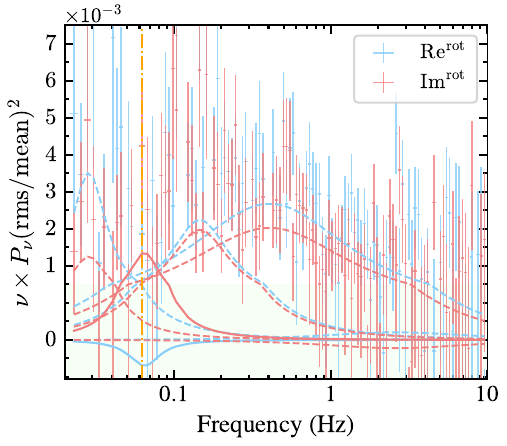}
            \caption{Cross spectrum of Fig.~\ref{Nicer_Example}.}
            \label{fig:linear_nicer}
        \end{subfigure}\hfill
        \begin{subfigure}[t]{0.49\linewidth}
            \vspace{0pt}
            \centering
            \includegraphics[width=\linewidth]{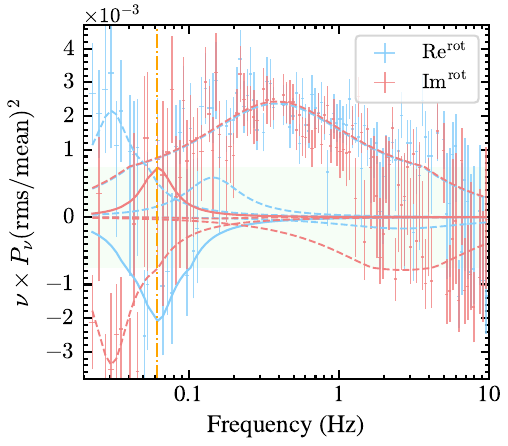}
            \caption{Cross spectrum of Fig.~\ref{HE_Example}.}
            \label{fig:linear_5C}
        \end{subfigure}
    \end{minipage}\hfill
    \begin{minipage}[c]{5.5cm}
        \vspace{0pt}
        \caption{Linear-scale cross spectra from Fig.~\ref{Nicer_Example} (left) and Fig.~\ref{HE_Example} (right). Solid and dashed curves denote the QPO and individual non-QPO components, respectively. In both panels, data with absolute values $\leq 5\times10^{-4}$ (light-blue region) are magnified by a factor of 3.}
        \label{fig:CS_linear}
    \end{minipage}
\end{figure*}

\section{Energy dependence of lag and coherence across the full frequency range}
\begin{figure*}[htbp]
    \centering
    \begin{minipage}[c]{12cm}
        \centering
        \includegraphics[width=\linewidth]{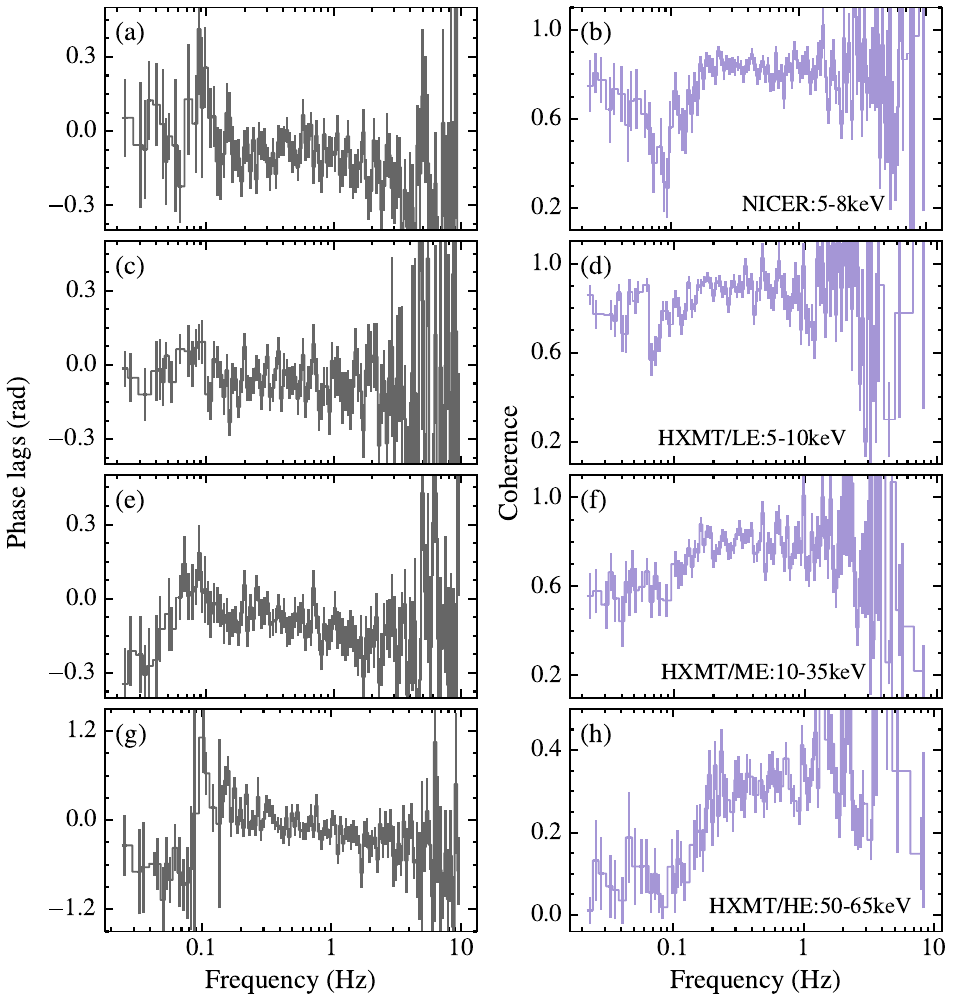}
    \end{minipage}\hfill
    \begin{minipage}[c]{5.5cm}
        \caption{Frequency-resolved phase lags (left) and coherence (right) of 1A~0535+262 on Day~7. Reference bands: 0.2--2\,keV (\textit{NICER}) and 1--3\,keV (HXMT/LE). Subject bands (top to bottom): \textit{NICER} 5--8\,keV, and HXMT/LE:5--10\,keV, HXMT/ME:27--35\,keV, and HXMT/HE:50--65\,keV. A transition from symmetric, modest bumps to sharp ``cliff'' structures is evident as energy increases, indicating the emergence of more pronounced QPO signatures in higher-energy bands.}
        \label{Cohlag_noFit}
    \end{minipage}
\end{figure*}
\begin{figure*}[htbp]
    \centering
    \begin{minipage}[c]{12cm}
        \centering
        \includegraphics[width=\linewidth]{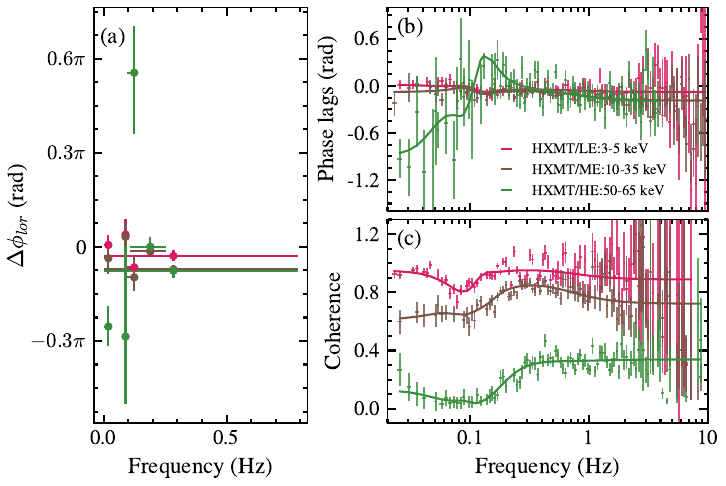}
    \end{minipage}\hfill
    \begin{minipage}[c]{5.5cm}
        \caption{Panel~(a): phase lag of the fitted Lorentzian components ($\Delta\phi_{\rm lor}$) versus frequency for three subject bands (HXMT/LE 3--5\,keV, HXMT/ME 10--35\,keV, and HXMT/HE 50--65\,keV; colours as labelled), measured relative to the HXMT/LE 1--3\,keV reference band. Panels~(b) and (c): frequency-resolved phase lags and coherence, respectively, for the same bands; solid curves show the best-fit models.}
        \label{Cohlag_Ene}
    \end{minipage}
\end{figure*}

Fig.~\ref{Cohlag_noFit} shows the energy evolution of the coherence and phase lags for four subject bands (\textit{NICER} 5–8\,keV, HXMT/LE 5–10\,keV, HXMT/ME 27–35\,keV, HXMT/HE 50–65\,keV; Day~7 at the outburst peak). 

In the softest band, the coherence remains near unity and exhibits only a shallow, symmetric dip at the QPO frequency. The corresponding phase lags display a mild, symmetric bump. Toward higher energies the dip broadens and deepens—approaching zero near the QPO frequency in the HE band—while the phase lags evolve into a sharp, asymmetric shape, crossing from $\lesssim\!-1$\,rad at low frequencies to positive values at the centroid frequency. 

Loss of coherence in X-ray binaries can result from non-linear spectral responses (e.g., ionization-dependent absorption and emission, thermal reverberation), spectral pivoting, and variability from spatially distinct or causally disconnected emission zones  \citep[e.g.][]{Nowak1999a,Ingram2009,Uttley2014,Grinberg2014,DeMarco2020,Mastroserio2021}. 

We attribute the observed behaviour to interference between multiple variability components present in the PS. As shown in the left panel of Fig.~\ref{Cohlag_Ene}, the broad-band components display different energy-dependent phase lags. At low energies (red), these components exhibit similar phase lags and high coherence. At higher energies (brown and green), however, their phase offsets begin to diverge, which reduces coherence around the QPO and gives rise to the sharp, cliff-like lag features seen at higher energies.

The above discussion can be formulated in the framework of \citet{Vaughan1997}, in which the observed light curves are modelled as the sum of multiple variability components contributing simultaneously to both energy bands, $x(t) = q_1(t) + r_1(t)$ and $y(t) = q_2(t) + r_2(t)$, where $q_1(t)$ and $q_2(t)$ represent the contributions of the first component in the reference and subject bands, and $r_1(t)$ and $r_2(t)$ are those of the second component in the same bands.
Let $Q_1$ and $Q_2$ denote the Fourier amplitudes of the first component in the two bands, and $R_1$, $R_2$ those of the second component.  
The phase lags of the two components between the bands are $\delta \theta_q$ and $\delta \theta_r$, respectively.

The intrinsic coherence is given by:
\begin{equation}
    \gamma_{xy}^2 = \frac{Q_1^2 Q_2^2 + R_1^2 R_2^2 + 2 |Q_1| |Q_2| |R_1| |R_2| \cos(\delta \theta_r - \delta \theta_q)}{Q_1^2 Q_2^2 + R_1^2 R_2^2 + Q_1^2 R_2^2 + Q_2^2 R_1^2}.
\end{equation}

When the two components are nearly out of phase (i.e., $\delta \theta_r - \delta \theta_q \approx {\pi\,{\rm rad}}$), the last term in the numerator becomes negative, leading to significant suppression of $\gamma_{xy}^2$.  
In our case, this condition—corresponding to the largest phase lag difference—is most strongly satisfied by the high-energy bands, consistent with the observed spectral features. These findings suggest that, in 1A~0535+262, the energy dependence of both the lag and coherence spectra is primarily governed by interference between multiple quasi-periodic and broadband components, with phase coupling strongly modulated by energy.

\section{Regrouped observations}

\begin{longtable}{cccccc}
\caption{\textit{NICER} and \textit{Insight}-HXMT observation log and QPO parameters for 1A~0535+262}
\label{tab:RegroupHXMT}\\
\toprule
Regroup-ID & NICER-ID & HXMT-ID & Start time (MJD) & QPO Frequency (mHz) & Luminosity ($10^{37}$ erg/s) \\
\midrule
\endfirsthead                                      
\toprule
Regroup-ID & NICER-ID & HXMT-ID & Start time (MJD) & QPO Frequency (mHz) & Luminosity ($10^{37}$ erg/s) \\
\midrule
\endhead
\midrule
\endfoot
\bottomrule
\endlastfoot                                           
\multirow{4}{*}{Day1} & \multirow{4}{*}{3200360130} & P031431600101 & 59167.33 & \multirow{4}{*}{$62.0^{+4.4}_{-3.8}$} & \multirow{4}{*}{$7.48$} \\
  & & P031431600102 & 59167.49 & & \\
  & & P031431600103 & 59167.63 & & \\
  & & P031431600105 & 59167.89 & & \\
\midrule
\multirow{7}{*}{Day2} & \multirow{7}{*}{3200360131} & P031431600106 & 59168.03 & \multirow{7}{*}{$74.0^{+3.2}_{-3.5}$} & \multirow{7}{*}{$9.06$} \\
  & & P031431600107 & 59168.16 & & \\
  & & P031431600108 & 59168.29 & & \\
  & & P031431600109 & 59168.42 & & \\
  & & P031431600110 & 59168.56 & & \\
  & & P031431600112 & 59168.82 & & \\
  & & P031431600113 & 59168.95 & & \\
\midrule
\multirow{5}{*}{Day3} & \multirow{5}{*}{3200360132} & P031431600202 & 59169.40 & \multirow{5}{*}{$88.0^{+2.4}_{-2.4}$} & \multirow{5}{*}{$10.78$} \\
  & & P031431600203 & 59169.55 & & \\
  & & P031431600204 & 59169.68 & & \\
  & & P031431600205 & 59169.82 & & \\
  & & P031431600206 & 59169.95 & & \\
\midrule
\multirow{6}{*}{Day4} & \multirow{6}{*}{3200360133} & P031431600207 & 59170.08 & \multirow{6}{*}{$89.0^{+3.2}_{-3.5}$} & \multirow{6}{*}{$11.61$} \\
  & & P031431600209 & 59170.35 & & \\
  & & P031431600210 & 59170.47 & & \\
  & & P031431600211 & 59170.61 & & \\
  & & P031431600212 & 59170.74 & & \\
  & & P031431600213 & 59170.88 & & \\
\midrule
     &            & P031431600214 & 59171.01 &                      &         \\
     &            & P031431600301 & 59171.30 &                      &         \\
     &            & P031431600302 & 59171.47 &                      &         \\
Day5 & 3200360134 & P031431600303 & 59171.60 & $86.4^{+2.4}_{-2.3}$ & $12.42$ \\
     &            & P031431600304 & 59171.74 &                      &         \\
     &            & P031431600305 & 59171.87 &                      &         \\
\midrule
\multirow{6}{*}{Day6} & \multirow{6}{*}{3200360135} & P031431600306 & 59172.00 & \multirow{6}{*}{$87.8^{+2.1}_{-2.4}$} & \multirow{6}{*}{$12.18$} \\
  & & P031431600309 & 59172.40 & & \\
  & & P031431600310 & 59172.53 & & \\
  & & P031431600311 & 59172.66 & & \\
  & & P031431600312 & 59172.80 & & \\
  & & P031431600313 & 59172.93 & & \\
\midrule
\multirow{5}{*}{Day7} & \multirow{5}{*}{3200360136} & P031431600314 & 59173.06 & \multirow{5}{*}{$91.0^{+2.6}_{-2.9}$} & \multirow{5}{*}{$12.52$} \\
  & & P031431600316 & 59173.33 & & \\
  & & P031431600317 & 59173.45 & & \\
  & & P031431600319 & 59173.72 & & \\
  & & P031431600320 & 59173.86 & & \\
\midrule
     &            & P031431600401 & 59174.22 &                      &         \\
     &            & P031431600402 & 59174.37 &                      &         \\
Day8 & 3200360137 & P031431600403 & 59174.52 & $84.9^{+2.4}_{-2.3}$ & $11.80$ \\
     &            & P031431600404 & 59174.65 &                      &         \\
     &            & P031431600405 & 59174.78 &                      &         \\
     &            & P031431600406 & 59174.92 &                      &         \\
\midrule
\multirow{8}{*}{Day9} & \multirow{8}{*}{3200360138} & P031431600407 & 59175.05 & \multirow{8}{*}{$85.1^{+2.2}_{-2.3}$} & \multirow{8}{*}{$11.88$} \\
  & & P031431600408 & 59175.18 & & \\
  & & P031431600409 & 59175.31 & & \\
  & & P031431600410 & 59175.44 & & \\
  & & P031431600411 & 59175.58 & & \\
  & & P031431600412 & 59175.71 & & \\
  & & P031431600413 & 59175.84 & & \\
  & & P031431600414 & 59175.98 & & \\
\midrule
\multirow{1}{*}{Day10} & \multirow{1}{*}{3200360139} & / & / & \multirow{1}{*}{$81.2^{+2.6}_{-4.5}$} & \multirow{1}{*}{$11.49$} \\
\midrule
\multirow{3}{*}{Day11} & \multirow{3}{*}{3200360140} & P031431600504 & 59177.70 & \multirow{3}{*}{$82.2^{+4.2}_{-4.2}$} & \multirow{3}{*}{$11.10$} \\
  & & P031431600505 & 59177.83 & & \\
  & & P031431600506 & 59177.96 & & \\
\midrule
\multirow{5}{*}{Day12} & \multirow{5}{*}{/} & P031431600508 & 59178.23 & \multirow{5}{*}{$82.2^{+0.3}_{-0.1}$} & \multirow{5}{*}{$10.89$} \\
  & & P031431600509 & 59178.36 & & \\
  & & P031431600511 & 59178.63 & & \\
  & & P031431600512 & 59178.76 & & \\
  & & P031431600513 & 59178.89 & & \\
\midrule
\multirow{7}{*}{Day13} & \multirow{7}{*}{3200360141} & P031431600515 & 59179.16 & \multirow{7}{*}{$82.1^{+2.0}_{-1.9}$} & \multirow{7}{*}{$10.61$} \\
  & & P031431600516 & 59179.29 & & \\
  & & P031431600517 & 59179.41 & & \\
  & & P031431600518 & 59179.55 & & \\
  & & P031431600519 & 59179.69 & & \\
  & & P031431600520 & 59179.82 & & \\
  & & P031431600521 & 59179.95 & & \\
\midrule
\multirow{5}{*}{Day14} & \multirow{5}{*}{3200360142} & P031431600601 & 59180.25 & \multirow{5}{*}{$80.9^{+1.0}_{-1.6}$} & \multirow{5}{*}{$10.16$} \\
  & & P031431600602 & 59180.41 & & \\
  & & P031431600603 & 59180.55 & & \\
  & & P031431600604 & 59180.68 & & \\
  & & P031431600605 & 59180.81 & & \\
\midrule
\multirow{6}{*}{Day15} & \multirow{6}{*}{3200360143} & P031431600608 & 59181.21 & \multirow{6}{*}{$77.7^{+2.0}_{-1.9}$} & \multirow{6}{*}{$9.53$} \\
  & & P031431600609 & 59181.34 & & \\
  & & P031431600610 & 59181.47 & & \\
  & & P031431600611 & 59181.61 & & \\
  & & P031431600612 & 59181.74 & & \\
  & & P031431600613 & 59181.87 & & \\
\midrule
      &            & P031431600615 & 59182.14 &                          &        \\
      &            & P031431600616 & 59182.27 &                          &        \\
      &            & P031431600617 & 59182.41 &                          &        \\
      &            & P031431600618 & 59182.53 &                          &        \\
Day16 & 3200360144 & P031431600619 & 59182.67 & $69.5^{+0.1}_{-0.9}$     & $8.80$ \\
      &            & P031431600620 & 59182.80 &                          &        \\
      &            & P031431600621 & 59182.94 &                          &        \\
\midrule
\multirow{5}{*}{Day17} & \multirow{5}{*}{3200360145} & P031431600801 & 59183.30 & \multirow{5}{*}{$69.3^{+0.9}_{-0.6}$} & \multirow{5}{*}{$7.92$} \\
  & & P031431600802 & 59183.46 & & \\
  & & P031431600803 & 59183.60 & & \\
  & & P031431600804 & 59183.73 & & \\
  & & P031431600805 & 59183.87 & & \\
\midrule
\multirow{6}{*}{Day18} & \multirow{6}{*}{/} & P031431600807 & 59184.13 & \multirow{6}{*}{$67.5^{+0.2}_{-0.2}$} & \multirow{6}{*}{$7.88$} \\
  & & P031431600808 & 59184.26 & & \\
  & & P031431600809 & 59184.40 & & \\
  & & P031431600810 & 59184.53 & & \\
  & & P031431600811 & 59184.66 & & \\
  & & P031431600812 & 59184.79 & & \\
      &            & P031431600814 & 59185.09 &                          &        \\
      &            & P031431600815 & 59185.24 &                          &        \\
      &            & P031431600816 & 59185.32 &                          &        \\
Day19 &    /       & P031431600817 & 59185.46 &   $66.3^{+1.1}_{-0.6}$   & $7.66$ \\
      &            & P031431600818 & 59185.59 &                          &        \\
      &            & P031431600819 & 59185.72 &                          &        \\
      &            & P031431600820 & 59185.86 &                          &        \\
\midrule
\multirow{3}{*}{Day20} & \multirow{3}{*}{3200360147} & P031431600901 & 59186.21 & \multirow{3}{*}{$64.2^{+0.1}_{-0.9}$} & \multirow{3}{*}{$6.92$} \\
  & & P031431600902 & 59186.37 & & \\
  & & P031431600903 & 59186.51 & & \\
\midrule
\multirow{3}{*}{Day21} & \multirow{3}{*}{3200360148} & P031431600905 & 59187.01 & \multirow{3}{*}{$58.5^{+1.2}_{-0.7}$} & \multirow{3}{*}{$6.27$} \\
  & & P031431600906 & 59187.15 & & \\
  & & P031431600907 & 59187.29 & & \\
\midrule
\multirow{5}{*}{Day22} & \multirow{5}{*}{3200360149} & P031431600910 & 59188.00 & \multirow{5}{*}{$56.5^{+1.7}_{-1.1}$} & \multirow{5}{*}{$5.67$} \\
  & & P031431600911 & 59188.14 & & \\
  & & P031431600912 & 59188.29 & & \\
  & & P031431600913 & 59188.42 & & \\
  & & P031431600914 & 59188.81 & & \\
\midrule
\multirow{4}{*}{Day23} & \multirow{4}{*}{3200360150} & P031431601001 & 59189.27 & \multirow{4}{*}{$53.7^{+0.4}_{-1.8}$} & \multirow{4}{*}{$5.23$} \\
  & & P031431601002 & 59189.42 & & \\
  & & P031431601003 & 59189.80 & & \\
  & & P031431601004 & 59189.92 & & \\
\midrule
\multirow{4}{*}{Day24} & \multirow{4}{*}{3200360151} & P031431601005 & 59190.06 & \multirow{4}{*}{$53.2^{+0.5}_{-2.9}$} & \multirow{4}{*}{$4.98$} \\
  & & P031431601006 & 59190.20 & & \\
  & & P031431601007 & 59190.34 & & \\
  & & P031431601008 & 59190.79 & & \\
  & & P031431601009 & 59190.92 & & \\
\midrule
\multirow{6}{*}{Day25} & \multirow{6}{*}{3200360152} & P031431601010 & 59191.05 & \multirow{6}{*}{$49.0^{+0.3}_{-0.1}$} & \multirow{6}{*}{$4.19$} \\
  & & P031431601011 & 59191.20 & & \\
  & & P031431601013 & 59191.50 & & \\
  & & P031431601014 & 59191.63 & & \\
  & & P031431601015 & 59191.76 & & \\
  & & P031431601016 & 59191.90 & & \\
\midrule
\multirow{5}{*}{Day26} & \multirow{5}{*}{3200360153} & P031431601101 & 59192.19 & \multirow{5}{*}{/} & \multirow{5}{*}{$3.70$} \\
  & & P031431601102 & 59192.33 & & \\
  & & P031431601103 & 59192.49 & & \\
  & & P031431601104 & 59192.62 & & \\
  & & P031431601105 & 59192.76 & & \\
\midrule
      &            & P031431601107 & 59193.02 &                          &        \\
      &            & P031431601108 & 59193.15 &                          &        \\
      &            & P031431601109 & 59193.29 &                          &        \\
      &            & P031431601110 & 59193.42 &                          &        \\
Day27 & 3200360154 & P031431601111 & 59193.55 & $44.1^{+3.9}_{-3.1}$     & $3.52$ \\
      &            & P031431601112 & 59193.68 &                          &        \\
      &            & P031431601113 & 59193.81 &                          &        \\
      &            & P031431601114 & 59193.95 &                          &        \\
\midrule
\multirow{6}{*}{Day28} & \multirow{6}{*}{3200360155} & P031431601115 & 59194.08 & \multirow{6}{*}{$40.6^{+2.0}_{-1.8}$} & \multirow{6}{*}{$3.20$} \\
  & & P031431601116 & 59194.21 & & \\
  & & P031431601117 & 59194.34 & & \\
  & & P031431601118 & 59194.48 & & \\
  & & P031431601119 & 59194.61 & & \\
  & & P031431601120 & 59194.74 & & \\
\end{longtable}
\noindent{Notes:} The QPO centroid frequencies are taken from fits in the 50--65\,keV band, except for Day~10, for which the \textit{NICER} 2--5\,keV band is used. Typical uncertainties on the luminosities listed in the last column are of order 1\%.

\clearpage

\end{appendix}
\end{document}